\documentclass[12pt, letterpaper, oneside]{article}

\usepackage{graphicx}
\usepackage[body={6.5in, 9in}, right=1in, top=1in]{geometry}
\usepackage{amssymb}
\usepackage{amsmath} 
\usepackage[linktoc=page,colorlinks=false,linkcolor=cyan,citecolor=cyan,urlcolor=blue]{hyperref}

\def\be{\begin{equation}}
\def\ee{\end{equation}}
\def\ba{\begin{eqnarray}}
\def\ea{\end{eqnarray}}	
\def\l{\left}
\def\r{\right}
\def\fr{\frac}
\def\la{\label}
\def\d{\partial}
\def\vphi{\varphi}

\def\mbf{\mathbf}

\begin{document}


\begin{center}
\LARGE{\textbf{A Modern Approach to Superradiance}} \\[1cm]
\large{Solomon Endlich$^{\rm a}$ and Riccardo Penco$^{\rm b, c}$}
\\[0.4cm]

\vspace{.2cm}
\small{\textit{$^{\rm a}$ Stanford Institute for Theoretical Physics, \\ Stanford University,
Stanford, CA 94306, USA}}

\vspace{.2cm}
\small{\textit{$^{\rm b}$ Physics Department, Center for Theoretical Physics\\
\& Institute for Strings, Cosmology, and Astroparticle Physics,\\
  Columbia University, New York, NY 10027, USA}}

\vspace{.2cm}
\small{\textit{$^{\rm c}$ Center for Particle Cosmology, Department of Physics and Astronomy, \\
University of Pennsylvania, Philadelphia, Pennsylvania 19104, USA}}

\end{center}

\vspace{.2cm}


\begin{abstract}
In this paper, we provide a simple and modern discussion of rotational superradiance based on quantum field theory. We work with an effective theory valid at scales much larger than the size of the spinning object responsible for superradiance. Within this framework, the probability of absorption by an object at rest completely determines the superradiant amplification rate when that same object is spinning. We first discuss in detail superradiant scattering of spin 0 particles with orbital angular momentum $\ell=1$, and then extend our analysis to higher values of orbital angular momentum and spin. Along the way, we provide a simple derivation of vacuum friction---a ``quantum torque'' acting on spinning objects in empty space. Our results apply not only to black holes but to arbitrary spinning objects. We also discuss superradiant instability due to formation of bound states and, as an illustration, we calculate the instability rate $\Gamma$ for bound states with massive spin 1 particles. For a black hole with mass $M$ and angular velocity $\Omega$, we find $\Gamma \sim (G M \mu)^7 \Omega$ when the particle's Compton wavelength $1/\mu$ is much greater than the size $GM$ of the spinning object. This rate is parametrically much larger than the instability rate for spin 0 particles, which scales like $(GM \mu)^9 \Omega$. This enhanced instability rate can be used to constrain the existence of ultralight particles beyond the Standard Model.
\end{abstract}

\newpage


\tableofcontents

\newpage


\section{Introduction}

Superradiance\footnote{Throughout this paper will we will continually use the more general term of ``superradiance'' in place of the more specific one ``rotational superradiance''.} is a surprising phenomenon where radiation interacting with a rotating object can be amplified if prepared in the correct angular momentum state~\cite{zel1971generation,zel1972amplification}. For an axially symmetric object, such amplification occurs whenever the following ``superradiant condition'' is met:
\be
\label{SR_condition}
\omega-m\Omega < 0 \, ,
\ee
where $\omega$ is the angular frequency of the incoming radiation, $m$ its angular momentum along the axis of rotation (which coincides with the axis of symmetry), and $\Omega$ is the magnitude of the angular velocity of the rotating object.

The importance of superradiance in astrophysics stems from the fact that it is a mechanism for extracting energy from spinning compact objects, and in particular from black holes~\cite{misner1972stability,starobinskii1973amplification,starobinskiichurilov1973amplification,teukolsky1974perturbations}. Because this rotational energy reservoir can be tremendous, any such mechanism could in principle have observable consequences and serve as a measure of strong gravity. Historically, however, it has been difficult to detect this phenomenon in real astrophysical systems. One main difficulty is that the amplification efficiency is generally very low\footnote{This is especially true in the long wavelength limit. However, it should be noted that the amplification factors for scalar and electromagnetic radiation are small at all frequencies ($<0.4\%$ and $<10\%$ respectively), whereas the amplification of gravitational radiation can become as high as $140\%$ at very high frequencies ($\omega$ of order the light crossing time)\cite{starobinskiichurilov1973amplification}.} for massless radiation~\cite{starobinskii1973amplification, starobinskiichurilov1973amplification,Press:1972zz}. This necessitates contrived scenarios such as the``Black Hole Bomb" of Press and Teukolsky~\cite{Press:1972zz}, where some sort of perfect spherical mirror encases the rotating object and reflects the amplified modes back allowing them to exponentially grow in energy. Consequently, it seems now that other astrophysical mechanisms, such as for instance the Blandford-Znajek process~\cite{Blandford:1977ds}, play a much more important role in the dynamics and evolution of compact objects than superradiant scattering of electromagnetic or gravitational radiation. 

Recently however, two distinct developments have led to a renewed interest in superradiance. First, the development of the gauge-gravity correspondence~\cite{Maldacena:1997re,Gubser:1998bc,Witten:1998qj} has spurred a great deal of activity surrounding black hole solutions in asymptotic AdS. For such solutions, the boundary of AdS acts as a perfect mirror reflecting gravitational radiation back to the black hole in a finite time making it possible for instabilities to develop~\cite{Cardoso:2004hs}. Secondly, in the context of particle physics it was realized that the existence of light particles beyond the standard model can affect the spin distribution of astrophysical black holes~\cite{Arvanitaki:2009fg}. Such light particles can become gravitationally bound to a black hole and, if they are bosons (such as axions), acquire extremely high occupation numbers. This instability can have a host of fascinating---and more importantly observable---consequences~\cite{Arvanitaki:2010sy, Arvanitaki:2014wva, Arvanitaki:2016qwi, Arvanitaki:2016fyj}.\footnote{We have given short shrift to a great deal of work on the subject matter of superradiance. The interested reader can find a much more complete record of the literature in the excellent reviews~\cite{Bekenstein:1998nt} and~\cite{Brito:2015oca}.}

These developments are representative of the fact that high energy physicists' interest in superradiance is often in the context of black holes. This may give the false impression that superradiance is somehow related to the existence of an ergosphere in the Kerr solution.\footnote{See for instance remarks to this end in the authoritative account of the Kerr metric by Teukolsky \cite{Teukolsky:2014vca}.} It is instead a much more general phenomenon that can occur for {\em any} rotating object that is capable of absorbing radiation. As a matter of fact, the original papers on the subject by Zel'dovich~\cite{zel1971generation,zel1972amplification}---beautiful in their brevity and clarity---are about scattering of electromagnetic radiation off a cylinder with finite conductivity. Nevertheless, discussions of superradiance are often obscured by the algebraic complexity of the Kerr solution. In this paper, we will show that, at least in the long wavelength limit, dealing with the details of the Kerr solution is neither necessary nor helpful. It is also very restrictive, because there is no analogue of Birkhoff's theorem for the Kerr solution~\cite{Teukolsky:2014vca}. This means that the metric outside a rotating star generically differs from the Kerr metric and can depend on additional parameters besides the mass $M$ and the spin $J$. It is therefore necessary to go beyond the Kerr solution in order to describe superradiant scattering off astrophysical objects other than black holes. This is however not feasible in the usual approach, which is based on finding solutions to the wave equation on a fixed curved background, because (\emph{a}) in general the exact form of the metric is not known, and (\emph{b}) even it was, the resulting wave equation would likely be much more complicated to solve analytically than for the Kerr metric.

The purpose of this work is to give a modern and comprehensive account of superradiance based on effective field theory (EFT) techniques, and to provide a simple framework to carry out perturbative calculations in the context of superradiant processes. By focusing on the long wavelength limit, our approach is capable of describing the onset of superradiance for any slowly rotating object, be that a star or a black hole.  This shows explicitly that superradiance is just a consequence of dissipation and spin, and nothing else. In particular, there is no need for an ergosphere. Moreover, it is possible to infer superradiant scattering efficiencies and superradiant instability rates by matching a single quantity (e.g. absorption cross section) which can be calculated \emph{even when the object is at rest}. For instance, one can extract the leading order results for rotating black holes from calculations carried out in a Schwarzschild background, without knowing anything about the Kerr metric.

The rest of this paper is organized as follows. In Section~\ref{EFT_for_spin}, we review the effective theory of relativistic spinning objects coupled to external fields discussed in~\cite{Delacretaz:2014oxa}. This approach is valid in the slowly-rotating regime, i.e. whenever the angular frequency is smaller than the object's characteristic frequencies.\footnote{For maximally rotating objects, the expansion in angular velocities breaks down~\cite{Delacretaz:2014oxa} and one needs to reorganize the effective action following for instance~\cite{Porto:2005ac,Levi:2015msa}.}  We also discuss how to incorporate the effects of dissipation---critical for superradiance---in a way that is consistent with unitarity and the EFT framework~\cite{Goldberger:2005cd,Porto:2007qi}. With the basic formalism in hand, we  illustrate our approach to superradiance in Section \ref{spin_0} by considering at first the $\ell =1$ modes of a spin $0$ field. Here, we argue that superradiant scattering follows from a tension between {\em absorption} and {\em stimulated emission}. Calculations are carried out in some detail, as similar manipulations take place also in the subsequent sections. In particular, we calculate the probabilities of absorption and spontaneous emission by considering processes that involve single quanta. In order to justify this approach and make contact with the standard one based on classical wave equations~\cite{Brito:2015oca}, in Section \ref{coherent states} we recalculate these probabilities using coherent states and find perfect agreement with the single-quantum approach in the limit of large occupation number.

We then generalize our results to higher values of the orbital angular momentum (Section \ref{higher multipoles}) and higher integer spins (Section \ref{higher spins}). These cases are a bit more cumbersome from a purely algebraic viewpoint, although conceptually they are simple generalizations of the results derived in Section~\ref{spin_0}. Finally, in Section \ref{bound states} we consider superradiant instability due to formation of bound states and show how to compute the instability rate using the formalism developed in the previous sections. For concreteness, we carry out explicit calculations for spin 0 and spin 1 particles. The latter case would be exceedingly complicated to analyze with conventional methods, because the wave equation for massive spin 1 fields is not factorizable on a Kerr background (let alone on more general axially symmetric backgrounds). Interestingly, we find the instability rate for spin 1 particles to be parametrically larger than the one for spin 0 particles, in agreement with the results of~\cite{Pani:2012bp}.  We conclude in Section~\ref{conclusions} by summarizing our results.  
\\

\noindent \emph{Conventions:} throughout this paper, we will work in units such that $c = \hbar =1$ and we will adopt a ``mostly plus'' metric signature. Greek indices $\mu$, $\nu$, $\cdots$ run over $0$, $1$, $2$, $3$ and capital Latin indices $I$, $J$, $\cdots$ run over $1$, $2$, $3$. Other conventions and technical details are summarized in the appendices.


\section{Spinning objects and dissipation}
\label{EFT_for_spin}

Spontaneous symmetry breaking is ubiquitous in Nature, and at macroscopic scales it becomes essentially unavoidable. In fact, any macroscopic object of finite-size is bound to break almost all space-time symmetries. It certainly breaks spatial translations by being in some place rather than somewhere else; it also breaks spatial rotations by having a particular orientation in space; finally, it breaks boosts by selecting a preferred reference frame---the one in which it is at rest. The only space-time symmetry that remains unbroken is time translations, as Newton's first law ensures that the speed of an object won't change over time unless it is acted upon by an external force. Note that spatial rotations are broken even if the object is highly symmetric, e.~g.~in the case of a perfect sphere. In this case, the symmetry of the object is encoded by an \emph{internal} symmetry group, which is spontaneously broken together with spatial rotations down to the diagonal subgroup~\cite{Delacretaz:2014oxa}. For simplicity, in this paper we will restrict ourselves to the case of spherically symmetric objects, and therefore our theory will have an unbroken diagonal $SO(3)$ symmetry. However, we should emphasize that this is by no means necessary.

At scales much larger than the size of the object, the most relevant degrees of freedom are the Goldstone modes associated with the symmetry breaking pattern described above. Although we are perhaps not accustomed to thinking of them in these terms, the Goldstones of translations are the spatial coordinates of the object and the Goldstones of rotations are its Euler angles~\cite{Delacretaz:2014oxa, Papenbrock:2013cra}. There are no Goldstones for the boosts, but this of course shouldn't be a cause for concern since the textbook version of Goldstone's theorem doesn't apply to space-time symmetries. As such, the number of Goldstones may not equal the number of broken space-time symmetries~\cite{Low:2001bw}.

The advantage of embracing this viewpoint is that one can rely on powerful techniques such as the coset construction~\cite{Callan:1969sn,Coleman:1969sm,Volkov:1973vd,ogievetsky:1974ab} to systematically write down an effective action for the Goldstone modes. Of course, one doesn't need the coset construction to know that the action for a relativistic point-like particle is
\begin{eqnarray} \la{pointlike particle action}
S = -m \int d \tau \sqrt{- \fr{d x^\mu}{d \tau} \fr{dx_\mu}{d \tau}} \, .
\end{eqnarray}
However, including the angular variables in a Lorentz-invariant way is less trivial~\cite{Hanson:1974qy,Balachandran:1979ha,Porto:2005ac}, as the Euler angles are not simply the spatial components of a 4-vector. The coset construction is one way\footnote{See \cite{Hanson:1974qy,Balachandran:1979ha,Porto:2005ac,Levi:2015msa} for alternative approaches.} to remove any guesswork from this procedure~\cite{Delacretaz:2014oxa}. Perhaps more importantly, from an effective theory viewpoint, one should generically include in the action all the terms compatible with the symmetries. Higher order corrections to the action (\ref{pointlike particle action}) are usually neglected only because they are suppressed by UV scales such as the size of the object~\cite{Goldberger:2004jt} or the frequencies of its normal modes~\cite{Delacretaz:2014oxa}. However, they describe observable phenomena such as tidal distortions due to gravity or elastic deformations due to centrifugal forces.  It is often important to take them into account and the coset construction provides a systematic way to do so.  

In a nutshell, given a symmetry breaking pattern, the output of the coset construction is a series of ``building blocks'' that depend on the Goldstones and belong to some representations of the unbroken symmetries---in our case, the diagonal $SO(3)$. By combing these building blocks in a way that preserves the \emph{unbroken} symmetries, we can write an action that is actually invariant under \emph{all} the symmetries, albeit with the spontaneously broken ones non-linearly realized and therefore not obviously manifest.  This action is organized as an expansion in powers of the angular velocities, and as such is only capable of describing slowly rotating objects. The exact form of the coset building blocks will not play any role in this paper, but we refer the interested reader to~\cite{Delacretaz:2014oxa} for more details.

The coset construction also provides us with instructions on how to couple additional fields to the Goldstone sector in a way that preserves the nonlinearly realized symmetries. This will be especially important in what follows, since we are ultimately interested in describing interactions between spinning objects and the long-wavelength (compared to the size of the object) modes of some field $\Phi^{\mu_1 ... \mu_n} $. According to the coset construction, we should first introduce the new field~\cite{Delacretaz:2014oxa} \footnote{For notational simplicity, we will not attempt to differentiate between indices in different frames (lab frame or co-rotating frame), as was done in~\cite{Delacretaz:2014oxa} and elsewhere.}
\begin{eqnarray}
\label{general_dressed_field}
\tilde \Phi^{\mu_1 ... \mu_n} \equiv  (\Lambda^{-1})^{\mu_1}{}_{\nu_1} \cdots (\Lambda^{-1})^{\mu_n}{}_{\nu_n}\Phi^{\nu_1 ... \nu_n}   \qquad \qquad\qquad  \Lambda^\mu{}_\nu \equiv B^\mu{}_\lambda (\beta^i) R^\lambda{}_\nu (\theta^i)  \, ,
\end{eqnarray}
where $B$ is a boost, $R$ a rotation, $\mbf \beta$ is the velocity of the spinning object (normalized to $c$) and the $\theta$'s are the Euler angles. The inverse Lorentz transformation $(\Lambda^{-1})^\mu{}_\nu$ essentially takes each component of $\Phi^{\mu_1 ... \mu_n}$ to the frame that is instantaneously comoving with the spinning object. Next, we should decompose the new tensor $\tilde \Phi^{\mu_1 ... \mu_n}$ into irreducible representations under the unbroken $SO(3)$. The transformation properties of these irreducible representations are such that rotationally invariant combinations built out of them and the building blocks for the Goldstones are guaranteed to also be Lorentz invariant. This allows us to write down Lorentz-invariant interactions that are localized on the worldline. Throughout this paper we will work in a reference frame where $\mbf \beta = 0$, so that the Lorentz transformation $\Lambda$ simply reduces to the rotation $R$. 

In order to describe dissipative phenomena such as superradiance it is necessary to introduce additional degrees of freedom besides the Goldstone modes. This is because dissipation is, by definition, a process in which energy and momentum are transferred from the large distance, macroscopic degrees of freedom (in our case, the Goldstones and the long-wavelength modes of the external field  $\Phi^{\mu_1 ... \mu_n}$) to the microscopic ones.  Following~\cite{Goldberger:2005cd,Porto:2007qi}, we are going to account for the latter ones in a model-independent way by introducing an infinite number of composite operators $\mathcal O^{I_1 \cdots I_n}$ that transform according to different representations of $SO(3)$.\footnote{See also~\cite{LopezNacir:2011kk},~\cite{Endlich:2015mke} and~\cite{Endlich:2012vt} for applications of this same method in cosmology, astrophysics and hydrodynamics respectively. Notice also that our approach to dissipation is essentially the point-like version of the Rytov formalism~\cite{rytov1989elements} employed in~\cite{Maghrebi:2012tv,Maghrebi:2012rp,Maghrebi:2014nca} to discuss spontaneous emission by rotating conductors at zero temperature and the resulting vacuum friction.} These operators are localized on the world-line and are defined in the rest frame of the object. As we will see, superradiance is a result of the interaction between external fields $\Phi^{\mu_1 ... \mu_n}$ and these  composite operators.


\section{Superradiant scattering} \la{spin_0}

In this section, we are going to discuss superradiance in the simplest possible case, namely that of a spherical wave of spin 0 particles with angular momentum $\ell=1$. We will generalize our analysis to higher integer spin particles and higher values of the orbital angular momentum in the following sections.


\subsection{Absorption} \la{absorption spin=0 l=1}

Let's start by studying the process where a spin 0 particle with frequency $\omega$ and angular momentum  $\ell, m$ is absorbed by a spinning object. Here, $m$ is the eigenvalue of the orbital angular momentum of the incoming particle along the axis of rotation of the object, which we will assume to be the $z$-axis. Schematically, we can denote this process as
\be \la{process}
X_i + (\omega, \ell, m) \to X_f  \,  .
\ee
where $X_i$ and $X_f$ are respectively the initial and final state of the spinning object. If we are not interested in the final state $X_f$, then the total probability $P$ for the absorption process can be obtained by summing over all possible final states: 
\be \la{P}
P_{\rm abs} = \sum_{X_f} \fr{|\langle X_f; 0| S | X_i; \omega, \ell, m \rangle |^2}{\langle \omega, \ell, m | \omega, \ell, m \rangle} \, ,
\ee
where we have assumed that the states $|X \rangle$ of the spinning object as well as the vacuum $| 0\rangle$ are normalized to 1, and the operator $S$ is given as usual by
\be \la{S}
S = T \exp \l\{ - i \int dt H_{\rm int} (t) \r\}  \, .
\ee
The probability $P_{\rm abs}$ can also be calculated in a slick way using the optical theorem~\cite{Goldberger:2005cd}, but here we will follow a more direct approach.

As explained in the previous section, the interaction Hamiltonian describing dissipative processes contains couplings between the fields interacting with the spinning object (in our case, a single scalar field $\phi$) and all possible composite operators $\mathcal{O}_{I_1, \cdots I_n}$. These operators should be thought of as ``living'' in the rest frame of the object. They encode all the microscopic degrees of freedom that we are not keeping track of and that are ultimately responsible for dissipation. In this section, we will only consider a coupling between $\phi$ and a composite operator that carries a single index, $\mathcal{O}_I$. Following from the more general form of Eq.~(\ref{general_dressed_field}), we can dress the scalar field so as to couple it to this composite operator. We will therefore work with the following interaction Hamiltonian:
\be \la{Hint spin 0}
H_{\rm int} = \d^I \phi \, R_I{}^J \mathcal{O}_J \, .
\ee
The field $\phi$ is evaluated at the location of the particle, which we will assume to be $\mbf x = 0$.

As we will see in Section \ref{higher multipoles}, a traceless composite operator with $\ell$ indices is responsible for superradiance of modes with angular momentum $\ell$. Hence, the interaction Hamiltonian (\ref{Hint spin 0}) will only allow us to discuss superradiance for modes with $\ell = 1$. This also explains why we are not considering the even simpler coupling with a composite operator $\mathcal{O}$ without any indices: it would only affect modes with $\ell=0$, for which there is no superradiance.

Let us focus first on the numerator that appears on the RHS of equation (\ref{P}). Using the fact that the states $|X_f \rangle$ form a complete set, we can rewrite it to first order in perturbation theory as
\ba \la{numerator 1}
\text{numerator} &\simeq& \sum_{X_f} \int dt dt'\langle X_i; \omega, \ell, m | H_{\rm int} (t')  |  X_f; 0 \rangle \langle X_f; 0| H_{\rm int} (t) | X_i; \omega, \ell, m \rangle  \\
&=&  \int dt dt'\langle \mathcal{O}_J(t') \mathcal{O}_L(t) \rangle\langle \omega, \ell, m | \d^I \phi (t') | 0 \rangle \langle 0 | \d^K \phi (t) |  \omega, \ell, m \rangle R_I{}^J (t') R_K{}^L (t) \, ,  \nonumber
\ea
where we have denoted with $\langle \cdots \rangle$ the expectation value on the initial state $|X_i \rangle$.
This expression can be further simplified by noting that, for spherically symmetric objects, the Wightman correlation function $\langle \mathcal{O}_J(t') \mathcal{O}_L(t) \rangle$ can only be proportional to $\delta_{JL}$. Then, its Fourier transform takes the form
\be
\langle \mathcal{O}_J(t') \mathcal{O}_L(t) \rangle = \delta_{JL}\int \fr{d \omega'}{2 \pi} \Delta (\omega') e^{i  \omega' (t-t')} \, ,
\ee
and the numerator (\ref{numerator 1}) reduces to
\be \la{numerator 2}
\text{numerator} =  \int \fr{d \omega'}{2 \pi} \Delta (\omega') \l| \int dt  e^{i  \omega' t} \langle 0 | \d^K \phi (t) |  \omega, \ell, m \rangle R_K{}^L (t) \r|^2 \,  .
\ee

It is in principle straightforward to calculate the quantity above, and therefore the probability $P_{\rm abs}$, using the decomposition of $\phi$ in terms of creation and annihilation operators. However, given that most of the results in this paper will follow from calculations very similar to this one, we find it worthwhile to provide some of the intermediate steps rather then simply quote the final result for $P_{\rm abs}$, which the impatient reader can find in eq. (\ref{final result absorption}).

We will first calculate the amplitude $\langle 0 | \d^K \phi (t) |  \omega, \ell, m \rangle$. Using the expression for the scalar product $ \langle \mbf{k} |  \omega, \ell, m \rangle$ given in the appendix,   and keeping in mind that $\phi$ is evaluated at $\mbf x = 0$, it is easy to show that
\ba
 \langle 0 | \d^K \phi (t) |  \omega, \ell, m \rangle &=&  \int \fr{d^3 k}{(2\pi)^3 \, 2 \omega_{k}} i k^K \langle 0 |  a_{k} |  \omega, \ell, m \rangle \, e^{- i \omega_{k} t} \nonumber \\
&=&  \int \fr{d^3 k}{(2\pi)^3 \, 2 \omega_{k}}  i k^K  \fr{\sqrt{2 \omega v}}{k}\,  (2 \pi)^2 \delta(\omega_k - \omega) Y_\ell^m (\hat{\mbf{k}})  e^{- i \omega_{k} t} \, . \la{amplitude 1}
\ea
We can further simplify this expression by rewriting the integral over $\mbf k$ in spherical coordinates. Then, the delta function takes care of the integral over the magnitude $k$, whereas the integral over the solid angle can be carried out by expressing $k^K$ in terms of spherical harmonics,
\be
\mbf{k} = k (\sin \theta \cos \vphi, \sin \theta \sin \vphi, \cos \theta) = k \sqrt{\fr{4 \pi}{3}} \l( - \fr{Y_1^{1 *}- Y_1^{-1 *}}{\sqrt{2}},  - i \, \fr{Y_1^{1 *}+ Y_1^{-1 *}}{\sqrt{2}}, Y_1^{0 *} \r) \, ,
\ee
and using the orthonormality property of the $Y_\ell^m$'s (see Appendix \ref{appa} for our normalization conventions). Eq. (\ref{amplitude 1}) then becomes
\ba \la{matrix element absorption l=1}
\langle 0 | \d_K \phi (t) |  \omega, \ell, m \rangle = \fr{i k^2}{\sqrt{6 \pi v \, \omega}} \, \delta_\ell^1 \, V_K^m  e^{- i \omega t} \, ,  
\ea
where $v = \d \omega / \d k$ is the group velocity of the incoming scalar particle, and 
\begin{eqnarray}\la{def V}
\mbf{V}^m  = \l(- \tfrac{1}{\sqrt{2}} (\delta_1^m - \delta_{-1}^m), -\tfrac{i}{\sqrt{2}} (\delta_1^m + \delta_{-1}^m), \delta_0^m \r) \, .
\end{eqnarray}
We have defined the complex vectors $\mbf{V}^m$ in such a way that they have unit norm, i.e. $V_n^L (V^m_L)^* = \delta^m_n$. In fact, these are just polarization vectors for states with angular momentum $\ell=1$, helicity $m$ and momentum in the $z$-direction. 

In order to carry out the integral over time in eq. (\ref{numerator 2}) we will need the explicit form of the rotation matrix $R_K{}^L (t)$, which for an object spinning around the $z$-axis with angular velocity $\Omega$ is simply
\be
 R_K{}^L (t)  = \l( \begin{array}{ccc} \cos (\Omega t) & - \sin  (\Omega t) & 0 \\ \sin  (\Omega t) & \cos  (\Omega t) & 0 \\ 0 & 0 & 1 \end{array}\r) \, .
\ee
It is then easy to check that 
\be \la{rotations of V around z}
V_K^m R^K{}_L (t) = V^m_L e^{i m \Omega t} \, .
\ee
After carrying out the integral over time, the numerator reduces to
\be
\text{numerator} = \fr{k^4\delta_1^\ell }{6 \pi \,  v \, \omega} \! \int \fr{d \omega'}{2 \pi} \Delta (\omega') \l| 2 \pi \delta ( \omega'-\omega + m \Omega)\r|^2 = \fr{k^4 \delta_1^l}{6 \pi \,  v \, \omega}\,  \Delta (\omega - m \Omega)  \, 2 \pi \delta (0) \, .
\ee
The singular factor of $2 \pi \delta (0)$ in the last expression is canceled by the denominator in eq. (\ref{P}), which with our choice for the normalization of states (see Appendix \ref{appa}) turns out to be precisely
\be
\text{denominator} = \langle \omega, \ell, m | \omega, \ell, m \rangle = 2 \pi \delta (0) \, .
\ee
We conclude therefore that the probability $P_{\rm abs}$ that a spin 0 particle with frequency $\omega$ and angular momentum  $\ell, m$ is absorbed due to the interaction with the composite operator $\mathcal{O}_I$~is 
\be \la{final result absorption}
P_{\rm abs} = \fr{k^4  \delta_1^\ell}{6 \pi \,  v \, \omega} \, \Delta (\omega - m \Omega) \, .
\ee

As advertised at the beginning of this calculation, the absorption probability due to the interaction with the composite operator $\mathcal{O}_I$ is non-zero only for modes with $\ell=1$. Notice also that $\Delta (\omega - m \Omega)$ is always real and positive\footnote{This is because, neglecting the indices for notational simplicity, $$\Delta (\omega) =\fr{\langle \mathcal{O} (\omega) \mathcal{O} (-\omega) \rangle}{2 \pi \delta (0)} = \fr{\langle |\mathcal{O} (\omega) |^2 \rangle}{\langle \omega, \ell, m | \omega, \ell, m \rangle} \geqslant 0,$$ where in the last step we used that  $\mathcal{O}(-\omega)=\mathcal{O}^*(\omega)$ because  $\mathcal O (t)$ is real.}, which means that we don't encounter any negative probability, unlike in the more traditional calculations \cite{Brito:2015oca, Porto:2007qi}. It is then straightforward to use this probability to calculate for instance the absorption cross section for incoming particles with angular momentum $\ell=1$, which is\footnote{See Appendix \ref{appb} for a derivation of the first equality.}
\be \la{cross section l=1}
\sigma (\omega, \ell=1, m) = \fr{3 \pi P_{\rm abs}}{k^2} = \fr{ k^2}{2 \,  v \, \omega} \, \Delta (\omega - m \Omega) \, .
\ee
%


\subsection{Spontaneous emission}

We will now show that the interaction Hamiltonian (\ref{Hint spin 0}) also leads to spontaneous emission of quanta of $\phi$. More precisely, we will now consider  the process
\be \la{Hawking-process}
X_i  \to X_f + (\omega, \ell, m) \, .
\ee
Once again we are not interested in the final state $X_f$ of the spinning object, and therefore the probability for such an event to occur is 
\be
P_{\rm em} = \sum_{X_f} \fr{|\langle X_f; \omega, \ell, m | S |  X_i; 0  \rangle |^2}{\langle \omega, \ell, m | \omega, \ell, m \rangle} \, ,
\ee
This probability is nearly identical to the absorption probability in equation (\ref{P}), and therefore we can use the same methods to calculate it. At lowest order in perturbation theory, we get
\ba
\text{numerator} &\simeq& \sum_{X_f} \int dt dt'\langle X_f; \omega, \ell, m | H_{\rm int} (t')  |  X_i; 0 \rangle \langle X_i; 0| H_{\rm int} (t) | X_f; \omega, \ell, m \rangle  \\
&=&  \int dt dt'\langle  \mathcal{O}_L(t) \mathcal{O}_J(t') \rangle\langle \omega, \ell, m | \d^I \phi (t') | 0 \rangle \langle 0 | \d^K \phi (t) |  \omega, \ell, m \rangle R_I{}^J (t') R_K{}^L (t)   \, .\nonumber 
\ea

Notice that the only difference compared to the numerator we calculated in the previous section is in the correlation function of the composite operators, which now is  $\langle \mathcal{O}_L(t) \mathcal{O}_J(t') \rangle$ whereas in eq. (\ref{numerator 1}) it was $\langle \mathcal{O}_J(t') \mathcal{O}_L(t) \rangle$. However, this correlation function is symmetric under $J \leftrightarrow L$ because of rotational invariance. It is therefore easy to see that the probability of spontaneous emission can be obtained from the absorption probability (\ref{final result absorption}) by replacing $\Delta (\omega-m \Omega)$ with $\Delta (m \Omega-\omega)$, i.e. 
\be \la{final result emission}
P_{\rm em} = \fr{k^4  \delta_1^\ell}{6 \pi \,  v \, \omega} \, \Delta (m \Omega-\omega) \, .
\ee
Once again, we stress that $\Delta (m \Omega-\omega)$ is always positive, and thus so is the probability $P_{\rm em}$.
It follows that the rate for spontaneous emission of spin 0 particles with $\ell=1$ angular momentum is\footnote{See Appendix \ref{appb} for a derivation of the first equality.}
\begin{eqnarray} \la{decay rate l=1}
\fr{d\Gamma (\omega, \ell=1, m)}{d \omega} =  \fr{k^2 P_{\rm em}}{2 \pi v \omega^2} = \fr{k^6}{12 \pi^2 \,  v^2 \, \omega^3} \,  \Delta (m \Omega-\omega) \, .
\end{eqnarray}

Before moving on to superradiance, let's pause for a moment to discuss the physical origin of such spontaneous emission of radiation. In particular, is this a thermal or a quantum phenomenon? It depends on the microscopic details of the spinning object. To illustrate this, let's first imagine that our dissipative spinning object is at zero temperature. Then, all the ``internal'' degrees of freedom that are modeled by our composite operators are in the ground state when viewed in the reference frame comoving with the object. Wightman correlation functions on the ground state are particularly simple, in that they are equal to the spectral density for positive frequencies and they vanish for negative frequencies~\cite{Arteaga:2008ux},~i.e.
\begin{eqnarray} \la{Delta at T=0}
\Delta (\omega) = \theta (\omega) \rho (\omega) \, .
\end{eqnarray}
Thus, according to eq.  (\ref{final result emission}) the probability of spontaneous emission is still non-zero even at zero temperatures, but only for modes s.t. $\omega - m \Omega <0$. The radiation emitted carries away angular momentum, and therefore the spinning object slows down over time and eventually comes to rest. This is a quantum phenomenon known as \emph{vacuum friction}\footnote{A similar phenomenon occurs also in media that otherwise are supposed to be frictionless, such as superfluids~\cite{Calogeracos:2001pw}. In this case, the particles that are spontaneously emitted are collective excitations, i.e. phonons and rotons.}. It was first conjectured by Zel'dovich~\cite{zel1971generation} and later discussed in much more detail in~\cite{Maghrebi:2012tv,Maghrebi:2012rp,Maghrebi:2014nca, manjavacas2010vacuum,manjavacas2010thermal}. The quantum nature of this phenomenon is also reflected by the fact that, in this case, the microscopic origin of dissipation is tied to quantum fluctuations---think for instance of a metallic object at zero temperatures, in which quantum fluctuations are responsible for exciting electrons within the conduction band leading to a non-zero conductivity.

We can also consider the complementary situation in which the object is at finite temperature but it is not spinning. Notice that this is not in contradiction with our definition of correlation functions as expectation values on the pure state $| X_i \rangle$. In fact, finite-temperature systems can also be described by a pure state provided one enlarges the Hilbert space~\cite{takahashi1996thermo}. Then, finite temperature corrections turn the step function in (\ref{Delta at T=0}) into a smooth slope, and the Wightman correlation function takes the form~\cite{Bellac:2011kqa}:
\begin{eqnarray} \la{Delta thermal}
\Delta (\omega) = [1 + n_B(\omega) ] \rho (\omega) \, ,
\end{eqnarray}
where $n_B(\omega) = (e^{\omega / T} -1)^{-1}$ is the familiar Bose distribution. In this case, the spontaneous emission occurs even for $\omega - m \Omega>0$ and is a thermal effect.


\subsection{Superradiant scattering} \la{sec:superradiance}

After calculating the absorption and emission probabilities, we are finally in a position to discuss superradiance. Let's consider therefore an incoming flux of particles $\Phi_{\rm in}$ with frequency $\omega$ and angular momentum $\ell=1$ but arbitrary $m$. After interacting with the spinning object, the net outgoing flux is determined by the competition between absorption and stimulated emission. When considered separately, the first one would yield an outgoing flux equal to $\Phi_{\rm in} (1- P_{\rm abs})$, whereas the latter one would give\footnote{Recall that the probability of stimulated emission is proportional to the number of ``spectator'' quanta.}  $\Phi_{\rm in} (1+ P_{\rm em})$. Combining these two effects, we find that the total outgoing flux is $\Phi_{\rm out} = \Phi_{\rm in} (1+ P_{\rm em}- P_{\rm abs})$ or, equivalently, that the relative change in flux is
\begin{eqnarray} \la{flux change}
\fr{\Phi_{\rm out}- \Phi_{\rm in}}{\Phi_{\rm in}} = P_{\rm em} - P_{\rm abs} = - \fr{k^4}{6 \pi \,  v \, \omega} \, \rho (\omega - m \Omega), \qquad \quad (\ell=1) \, ,
\end{eqnarray}
where we used the fact that, by definition, the difference $\Delta (\omega) - \Delta (- \omega)$ is equal to the spectral density $\rho (\omega)$~\cite{Arteaga:2008ux}. Superradiance is then just a consequence of the fact that the spectral density of bosonic operators is an odd function of its argument, and that it is positive for positive arguments.\footnote{To avoid any potential confusion, we should point out that there are two notions of spectral density in the literature. In this paper, we are following the conventions of~\cite{Bellac:2011kqa}. The combination $ \text{sign} (\omega) \rho(\omega)$ is also sometimes referred to as spectral density (see for instance~\cite{Weinberg:1995mt}), but this combination is instead a positive definite, even function of $\omega$.  We refer the reader to~\cite{Arteaga:2008ux} for a pedagogical discussion of spectral densities in the vacuum as well as in more general states.} These properties are completely general~\cite{Arteaga:2008ux} and do not rely on any assumption about the composite operators nor the state of the spinning object. It follows therefore from eq. (\ref{flux change}) that $\Phi_{\rm out} > \Phi_{\rm in}$ ($\Phi_{\rm out} < \Phi_{\rm in}$) when $\omega - m \Omega < 0$ ($\omega - m \Omega > 0$).

\subsection{Matching at low energies}

The results derived thus far take an even simpler form if we are willing to make some physical assumptions about the composite operators $\mathcal O$ and the initial state $| X_i \rangle$ of the spinning object. Correlation functions of quantities that are not conserved charges---such as our composite operators $\mathcal O$---must decay faster than any power at large times in thermalized systems~\cite{Endlich:2012vt}. This means that their spectral density must admit a low-frequency Taylor expansion around $\omega =0$. Moreover, because the spectral density of bosonic operators is an odd function of $\omega$ that is positive for $\omega >0$, we can approximate it at low-energies as 
\begin{eqnarray} \la{rho small omega}
\rho (\omega) \simeq \gamma \, \omega + O(\omega^3), \qquad \quad \text{with} \qquad\quad  \gamma > 0 \, ,
\end{eqnarray}
where $\gamma$ is a coefficient that can depend on the temperature but, importantly, not on the spin of the object since the operators $\mathcal{O}$ ``live'' in the rest frame. Therefore, $\gamma$ can in principle be extracted from numerical simulations or analytical calculations in a simple, idealized problem (say, by calculating the absorption cross-section for a massless particle when the object is at rest) and then brought to bear on more complicated processes (say, superradiant scattering of a massive particle when the object is spinning).\footnote{For higher spins, one cannot immediately infer the properties of massive particles from those of massless ones as the number of degrees of freedom changes discontinuously and necessitates a more careful treatment. We will discuss this more at length in Sec.~\ref{higher spins}.}

As an example, let's see how this works in the case of black holes. The absorption probability for a massless spin 0 particle with $\ell =1$ from a Schwarzschild black hole in the long wavelength limit is~\cite{Page:1976df}
\begin{eqnarray}
P_{\rm abs} = \frac{r_s^4 \omega^4}{9} \,  ,
\end{eqnarray}
with $r_s = 2 GM$ the Schwarzschild radius. This result should be matched with the one in eq. (\ref{final result absorption}). From a classical viewpoint, a black hole should be regarded as an object in its ground state, since Hawking emission is a purely quantum mechanical effect. Therefore, the correlation function $\Delta (\omega)$ takes the form in (\ref{Delta at T=0}), and using the low-frequency expansion of the spectral density in eq. (\ref{rho small omega}) we find that $\gamma = \frac{2}{3} \pi r_s^4$ for a Schwarzschild black hole. Moreover, with this identification the absorption cross section in (\ref{cross section l=1}) reproduces the one calculated in~\cite{Unruh:1976fm} for massive spin 0 particles when $\Omega = 0$. 

We can now plug this value of $\gamma$ in eq. (\ref{flux change}) to predict the relative change in the intensity of a beam of massive spin 0 particles due to superradiant scattering from a Kerr black hole:
\begin{eqnarray}
\Delta \Phi \simeq - \fr{r_s^4 k^4  (\omega - m \Omega)}{9 \,  v \, \omega} \, .
\end{eqnarray}
Note that this result is just the leading term in an expansion in small $\Omega$ (one can of course account for higher order terms in a systematic way by including higher order corrections in the effective action).  In the massless limit where $v \to 1$ and $ k \to \omega$, our result reduces indeed to the known result in the literature~\cite{starobinskii1973amplification}.


\section{Coherent states} \la{coherent states}

Before extending our analysis to higher multipoles and higher spins, we will pause for a moment and discuss the connection between our derivation of superradiance, based on absorption and emission of single quanta, and earlier derivations of superradiance based on solutions to classical wave equations. From a quantum field theory perspective, classical waves are states with a very large occupation number. As such, the closest analog to the classical understanding of superradiance would be the computation of the transition amplitude from one state with large occupation number to another. This calculation can be carried out explicitly using coherent states and, as we will see, the final result is the same as the one derived by considering single quanta.

To be more precise, we are going to compute the probability for the process
\be \la{process_coherent}
X_i +  \alpha \to X_f + \beta  \, ,
\ee
where $\alpha$ and $\beta$ are coherent states for a scalar field in the basis of spherical waves, i.e. they are eigenstates of the lowering operators $a_{\omega \ell m}$:
\be \label{coherent_state_relation}
a_{\omega \ell m} | \alpha \rangle = \alpha_{\omega \ell m}| \alpha \rangle  \,  ,
\ee
and similarly for $|\beta\rangle$. Explicitly, a coherent state of a free scalar in the basis of spherical waves is given by
\be
|  \alpha \rangle = N_\alpha \exp \left(  \sum_{\ell m}\int \frac{d \omega}{2\pi} \, \alpha_{\omega \ell m} a^\dagger_{\omega \ell m} \right)| 0 \rangle  \,  ,
\ee
where $N_\alpha=\exp \left(-\frac{1}{2} \int_\omega \sum_{\ell m} |\alpha|^2 \right)$ such that $| \alpha \rangle$ is normalized to one. It is easy to check explicitly that such a state indeed satisfies eq.~(\ref{coherent_state_relation}). Moreover, using this definition one can also show that the inner product of two different coherent states is equal to
\be
\label{inner_prod}
\langle  \beta | \alpha \rangle=\exp \left(-\frac{1}{2} \sum_{\ell m} \int \frac{d\omega}{2\pi} \, |\alpha_{\omega \ell m} -\beta_{\omega \ell m} |^2 \right)  \,  .
\ee

Again, since we are not interested in observing the final state of the spinning object $X_f$, we are going to sum over all possible final states. The probability for this process to occur is then given by
\be \la{P_coherent}
P_{\alpha \to \beta} = \sum_{X_f} |\langle X_f;  \beta | S | X_i; \alpha \rangle |^2  \,  ,
\ee
as we have chosen the coherent states to be normalized to $1$. 
Expanding to leading order in the interactions we obtain
\ba
 P_{\alpha \to \beta} &\simeq&\int dt dt'\langle \alpha | \d^I\phi R_I{}^J(t')  |\beta \rangle \langle \beta | \d^K\phi R_K{}^L(t)  |  \alpha \rangle \langle X_i | \mathcal{O}_J(t')   \mathcal{O}_L(t) |  X_i \rangle \label{coherent_state_full} \\
&& - \frac{1}{2!} \l\{ \int dt dt' \langle \alpha | \beta \rangle  T \bigg[ \langle  \beta|  \d^I\phi R_I{}^J(t') \d^K\phi R_K{}^L(t)   |  \alpha \rangle  \langle X_i |\mathcal{O}_J(t')   \mathcal{O}_L(t) |  X_i \rangle\bigg] +\text{c.c.} \r\} \, .   \nonumber
\ea
The term in the first line is a generalization of the term we have already calculated in the case of a single quantum, whereas the one in the second line is new. The latter arises because different coherent states are not necessarily orthogonal---see eq. (\ref{inner_prod})---while in the case of a single quantum we have that $\langle \omega, \ell, m | 0 \rangle = 0$. Let us consider these two terms separately.

When it comes to the first term, we can use manipulations similar to those employed in section \ref{absorption spin=0 l=1} to show that
\ba
\langle \beta | \d^K\phi R_K{}^L(t)  | \alpha \rangle = \langle \beta  | \alpha \rangle \sum_m  \int_M^\infty \frac{d\omega}{2\pi} \frac{k^2}{\sqrt {6\pi v \omega}}\! \left[\alpha_{\omega 1m}V_m^Le^{-i(\omega-m\Omega)t}+ \beta_{\omega1m}^*(V_m^L)^*e^{i(\omega-m\Omega)t}\right]
\ea
where we have used the defining relation of coherent states, eq. (\ref{coherent_state_relation}), and the triplet of vectors $V_m^L$ was defined in eq. (\ref{def V}). Using this result, we can rewrite the first line of eq. (\ref{coherent_state_full})~as
\ba
&&\int dt dt'\langle \alpha | \d^I\phi R_I{}^J(t')  |  \beta \rangle \langle \beta | \d^K\phi R_K{}^L(t)  |\alpha \rangle \Delta_{JL}(t'-t) = \nonumber \\ 
&& \qquad \qquad =  |\langle \alpha  | \beta \rangle |^2\sum_m \int_M^\infty \frac{d\omega}{2\pi}\frac{k^4}{6\pi v \omega}\, \bigg[ | \alpha_{\omega 1m} |^2 {\Delta}(\omega-m\Omega)+| \beta_{\omega 1m} |^2 {\Delta}(m\Omega-\omega)\bigg]  \, .\la{first term rewritten} \qquad 
\ea
In performing the above computation we were able to drop the terms proportional to $ \alpha_{\omega 1 m} \beta_{\omega' 1 m}^*$ and its complex conjugate because they are proportional to $\delta(\omega+\omega')$ which vanishes everywhere on the domain of integration.

Continuing on with the second term in (\ref{coherent_state_full}) and its complex conjugate, the first order of business is dealing with the time ordered product. After expressing it in terms of theta functions, we can relabel the integration variables $t$ and $t'$ to rewrite the second line in eq. (\ref{coherent_state_full}) as follows:
\ba \la{second term}
- \int dt dt' \langle \alpha  | \beta \rangle  \theta (t-t') \langle  \beta  |  \d^I\phi R_I{}^J(t) \d^K\phi R_K{}^L(t')   | \alpha  \rangle \langle X_i |\mathcal{O}_J(t)   \mathcal{O}_L(t') |  X_i \rangle + \text{c.c.} 
\ea
Using the integral representation of the $\theta$ function
\be
\theta(t-t')=\int \frac{d \omega }{2\pi} \frac{e^{-i\omega(t-t')}}{\omega+i\varepsilon}\qquad \qquad\quad  (\varepsilon \rightarrow 0^+) \, ,
\ee
we can expand the scalar field into creation and annihilation operators and use again the defining properties of the coherent states to compute the expression in eq. (\ref{second term}). We find that the terms $\propto a^\dagger a^\dagger $ and $a a$ have vanishing support over the integration range, whereas the $a a^\dagger$ can be dealt with using the commutation relations. Formally, this procedure also yields a divergent contribution. However, this is simply a vacuum bubble that must be modded out to ensure that the final result remains finite and it reads 
\ba
-\left|\langle \alpha  | \beta \rangle \right|^2  \sum_m \int \frac{d \omega'}{2\pi} \int_M^\infty \frac{d\omega}{2\pi} \frac{k^4\beta_{\omega 1m}^* \alpha_{\omega 1m} }{6 \pi v \omega}  \left[\frac{i\Delta( \omega')}{(\omega-m\Omega)-\omega' +i\varepsilon} +\frac{i\Delta(\omega')}{(m\Omega-\omega)-\omega'+i\varepsilon}\right] +\text{c.c.} \nonumber 
\ea

At this point, the probability $P_{\alpha \to \beta}$ is equal to the sum of this result plus the one in eq. (\ref{first term rewritten}).
To make further progress, we need to make additional physical assumptions about the incoming and outgoing classical waves. First, we will assume that the two waves are in phase, which amounts to requiring that the coefficients $\alpha_{\omega 1m}$ and $\beta_{\omega1m}$ have the same phase. Since our probability depends only on the absolute values of these coefficients and on the products $\alpha \beta^*$ and $\beta \alpha^*$, the phase of the coefficients becomes irrelevant and from now on we will treat  $\alpha_{\omega 1m}$ and $\beta_{\omega1m}$ as if they were real. 

Second, we will assume that both incoming and outgoing waves are Gaussian wave-packets centered around the same frequency $\omega_\star$ with variance $\sigma$. One could in principle also consider two wave packets with different mean frequencies and variances. However, in order to make contact with the previous section we will eventually send the variance of these wave-packets to zero so that the waves become monochromatic. In this limit one finds that the probability $P_{\alpha \to \beta}$ vanishes unless the two waves have the same frequency. Therefore, we will assume from the very beginning that the coefficients $\alpha_{\omega 1m}$ and $\beta_{\omega1m}$ have the following form:
\begin{subequations} \la{alpha beta gaussian}
\ba 
\alpha_{\omega 1m}&=&\frac{\mathcal A_{m}(4\pi)^{1/4}}{ \sqrt{\sigma}} \exp \left(-\frac{(\omega-\omega_\star)^2}{2\sigma^2} \right)\\
\beta_{\omega 1m}&=&\frac{\mathcal B_{m}(4\pi)^{1/4}}{ \sqrt{\sigma}} \exp \left(-\frac{(\omega-\omega_\star)^2}{2\sigma^2} \right)
\ea
\end{subequations}
where the parameters $\mathcal A_{m} $ and $\mathcal B_{m} $ are constants that determine the waves' amplitude. Notice that, to further simplify matters, we have also assumed that the two wave-packets have the same variance $\sigma$, and added an overall factor $(4\pi)^{1/4}$ to the normalization. 

Plugging equations (\ref{alpha beta gaussian}) into our expression for $P_{\alpha \to \beta}$ and taking the limit $\sigma \to 0$, we can use the following representations of the delta function
\be
\delta(x)  = \lim_{\sigma \rightarrow 0} \, \frac{1}{\sigma \sqrt \pi}e^{-x^2/\sigma^2} = \lim_{\varepsilon \rightarrow 0} \, \fr{i}{2 \pi} \l[ \fr{1}{x+i \varepsilon} - \fr{1}{x-i \varepsilon}\r]
\ee
to write the probability $P_{\alpha \to \beta}$ as
\ba
P_{\alpha \to \beta} \!\!&=& \!\! \left|\langle \alpha  | \beta \rangle \right|^2  \sum_m \frac{k_\star^4}{6\pi v_\star \omega_\star} \bigg\{\mathcal A_{m}^2{\Delta}(\tilde\omega_\star)+\mathcal B_{m}^2{\Delta}(-\tilde\omega_\star) - \mathcal A_{m} \mathcal B_{m}\left[\Delta(\tilde\omega_\star)+\Delta(-\tilde\omega_\star)\right]\bigg\}  \\ 
&\simeq& \!\! \prod_m \exp \bigg\{- (\mathcal A_{m} - \mathcal B_{m})^2+ \mathcal{N}_\star \Big[ \mathcal A_{m}^2 \Delta(\tilde\omega)+\mathcal B_{m}^2 \Delta(-\tilde\omega)  - \mathcal A_{m} \mathcal B_{m}\left( \Delta(\tilde\omega)+\Delta(-\tilde\omega) \right)\Big]\bigg\} \nonumber 
\ea
where we have simplified the notation by defining $\tilde \omega_\star \equiv \omega_\star-m\Omega$ and $ \mathcal N_\star  \equiv k^4_\star/(6\pi v_\star \omega_\star)$ and, in the second step, we have used the result (\ref{inner_prod}) for the inner product of two coherent states. Our final expression is accurate at leading order in the interactions. 

In the classical limit, where the amplitude of the coherent states become very large, only the process for which the argument of the  exponential vanishes has non-negligible probability to occur. Setting the exponent equal to zero, we can solve for the outgoing amplitude $\mathcal B_{m}$ up to next-to-leading order in the interactions to find
\be
\mathcal B_{m} = \mathcal A_{m} \left[1-\frac{k_\star^4}{12\pi v_\star \omega_\star} \rho(\omega_\star-m\Omega) \right]  \,  ,
\ee
with the spectral density $\rho (\omega) = \Delta (\omega) - \Delta (- \omega)$. The intensity of the incoming (outgoing) wave with $\ell =1$ and arbitrary $m$ is proportional to $|\alpha_{\omega 1m}|^2$ ($|\beta_{\omega 1m}|^2$). Therefore, the relative change in the intensity is equal~to
\begin{eqnarray}
\fr{|\mathcal B_{m}|^2 - |\mathcal A_{m}|^2}{|\mathcal A_{m}|^2} \simeq -\frac{k_\star^4}{6\pi v_\star \omega_\star} \rho(\omega_\star-m\Omega), \qquad \quad (\ell=1) \, .
\end{eqnarray}
This is exactly the same result that we derived in the previous section by considering scattering of a single particle---see eq. (\ref{flux change}). Since our calculation based on coherent states supports the validity of our single-quantum approach, in the rest of this paper we are only going to consider processes that involve single quanta.


\section{Higher multipoles} \la{higher multipoles}

We are now going to repeat our analysis of absorption, emission and superradiance for higher values of the angular momentum $\ell$. Since the calculations are very similar to those we discussed in the previous section, we will only highlight the few differences. The dominant contribution will now come from the coupling with a composite operator with $\ell$ indices. We will therefore consider an interaction Hamiltonian of the form
\begin{eqnarray} \la{higher multipoles spin 0 interaction}
H_{\rm int} = \d^{I_1} \cdots \d^{I_\ell} \phi \, R_{I_1}{}^{J_1} \cdots  R_{I_\ell}{}^{J_\ell} \mathcal{O}_{J_1 \cdots  J_\ell} \, .
\end{eqnarray}
We can restrict ourselves to the case in which $\mathcal{O}$ is completely symmetric and traceless as we are interested only in the leading contribution to modes of a given angular momentum $\ell$. If we had included various traces, these would simply generate higher order corrections for modes of angular momentum $\ell-2n$, where $n$ is the number of traces. Consequently, the 2-point function of the $\mathcal{O}$ operators can be expressed in Fourier space as
\be \la{fourier 2-pt functions higher integer spin}
\langle \mathcal{O}^{I_1 \cdots  I_\ell}(t') \mathcal{O}_{J_1 \cdots  J_\ell}(t) \rangle = \delta^{I_1 \cdots  I_\ell}_{J_1 \cdots  J_\ell} \int \fr{d \omega}{2 \pi} \, \Delta_\ell (\omega) \, e^{i  \omega (t-t')} \, ,
\ee
where $\delta^{I_1 \cdots  I_\ell}_{J_1 \cdots  J_\ell}$ is the identity on the space of traceless, symmetric, rank-$\ell$ tensors.

Now, instead of the matrix element (\ref{matrix element absorption l=1}), we will be interested in the quantity
\ba \la{matrix element absorption l}
\langle 0 | \d_{I_1} \cdots \d_{I_\ell} \phi (t) |  \omega, \ell, m \rangle &=& \fr{(i)^\ell \sqrt{\ell!} \, k^{\ell+1} e^{- i \omega t} }{\sqrt{2 \pi v \omega (2\ell+1)!!}} V^{m}_{I_1 \cdots  I_\ell} \, ,  
\ea
The $V^{m}_{I_1 \cdots  I_\ell}$ are a higher-$\ell$ generalization of the quantity $V_I^m$ introduced in the previous section. For $-\ell \leqslant m \leqslant \ell$, they form a basis for the $(2\ell+1)$-dimensional vector space of rank-$\ell$ symmetric and traceless tensors, and they are equal to\footnote{The quantities $V^m_{I_1 \cdots  I_\ell}$ are related to the $\mathcal{Y}^{\ell m}_{I_1 \cdots  I_\ell}$ in~\cite{Thorne:1980ru} by $V^m_{I_1 \cdots  I_\ell} = \sqrt{\fr{4 \pi \ell!}{(2\ell+1)!!}} \mathcal{Y}^{\ell m}_{I_1 \cdots  I_\ell} $.}
\begin{eqnarray}
V^m_{I_1 \cdots  I_\ell} =  \sum_{j=0}^{[(\ell-m)/2]} \nu^{\ell mj} (\delta^1_{(I_1}+i \delta^2_{(I_1}) \cdots (\delta^1_{I_m}+i \delta^2_{I_m})\delta^3_{I_{m+1}} \cdots \delta^3_{I_{\ell-2j}} (\delta^{a_1}_{I_{\ell-2j+1}}\delta^{a_1}_{I_{\ell-2j+2}}) \cdots (\delta^{a_j}_{I_{\ell-1}}\delta^{a_j}_{I_{\ell)}}) \nonumber 
\end{eqnarray}
for $m \ge 0$, and where $[(\ell-m)/2]$ means ``the largest integer less than or equal to $(\ell-m)/2$'', and 
\begin{eqnarray}
\nu^{\ell mj} = \sqrt{\fr{\ell! (\ell-m)!}{(2\ell-1)!!(\ell+m)!}} \,  \fr{ (-1)^{m+j} (2\ell - 2j)!}{2^\ell j! (\ell-j)!(\ell-m-2j)!} \, .
\end{eqnarray}
For $m<0$, $V^m_{I_1 \cdots  I_\ell} \equiv (-)^m (V^{|m|}_{I_1 \cdots  I_\ell})^*$. We have chosen the overall normalization in such a way that $(V^m_{I_1 \cdots  I_\ell})^* V_{m'}^{I_1 \cdots  I_\ell} = \delta^m_{m'}$. Once again, the $V^m_{I_1 \cdots  I_\ell}$'s can be thought of as polarization tensors for states with angular momentum $\ell$, helicity $m$  and momentum in the $z$-direction.

It is now possible to show that the quantities $V^m_{I_1 \cdots  I_\ell}$ transform under rotations around the $z$-axis very much like in equation (\ref{rotations of V around z}), namely 
\begin{eqnarray} \la{property of V} 
V^m_{I_1 \cdots  I_\ell} R^{I_1}{}_{J_1} \cdots  R^{I_\ell}{}_{J_\ell} = V^m_{J_1 \cdots  J_\ell} e^{i m \Omega t} \, .
\end{eqnarray}
Using this result, one finds that the probability of absorption of a scalar particle with frequency $\omega$ and angular momentum $\ell,m$ is 
\begin{eqnarray} \la{absorption probability higher multipoles}
P_{\rm abs} = \fr{\ell! \, k^{2 \ell + 2}}{2 \pi (2 \ell+1)!! v \omega} \Delta_\ell (\omega -m \Omega) \, ,
\end{eqnarray}
the emission probability is
\begin{eqnarray} 
P_{\rm em} = \fr{\ell! \, k^{2 \ell + 2}}{2 \pi (2 \ell+1)!! v \omega} \Delta_\ell (m \Omega- \omega) \, ,
\end{eqnarray}
and the relative difference between outgoing and incoming flux is 
\begin{eqnarray} \la{flux change l}
\fr{\Phi_{\rm out}- \Phi_{\rm in}}{\Phi_{\rm in}} = P_{\rm em} - P_{\rm abs} = -  \fr{\ell! \, k^{2 \ell + 2}}{2 \pi (2 \ell+1)!! v \omega} \, \rho_\ell (\omega - m \Omega) \, .
\end{eqnarray}
Once again, the spectral densities $\rho_\ell$ are odd functions that at low energies can be approximated as $\rho_\ell (\omega) \simeq \gamma_\ell \, \omega$ with $\gamma_\ell >0$.  This shows that also higher multipoles are susceptible to superradiance. The values of the coefficients $\gamma_\ell$ appropriate for, say, a black hole can easily be calculated by matching the \emph{massless}, zero temperature limit of the absorption probability (\ref{absorption probability higher multipoles}) with the results in~\cite{Page:1976df}. The same values of $\gamma_\ell$ will describe also dissipative interactions between black holes and \emph{massive} spin 0 particles.


\section{Higher spins} \la{higher spins}

We are now in a position to extend the results discussed so far to higher integer spin particles. A field of arbitrary integer spin can be decomposed into the following sum of creation and annihilation operators:
\begin{eqnarray} \la{creation annihilation expansion arbitrary integer spin}
\hat{\Phi}_{\mu_1 \cdots \mu_s} (x) = \sum_{\lambda} \int \fr{d^3k}{(2 \pi)^3 2 \omega_k} \l\{  \hat{a}_{\mbf k, \lambda} \epsilon_{\mu_1, \cdots \mu_s} (\hat k, \lambda) e^{i k \cdot x} + \hat{a}_{\mbf k, \lambda}^\dag  \epsilon_{\mu_1, \cdots \mu_s}^* (\hat k, \lambda) e^{- i k \cdot x}  \r\} \, ,
\end{eqnarray}
where the polarization vectors $ \epsilon_{\mu_1, \cdots \mu_s} $ are normalized in such a way that 
\begin{eqnarray}
\epsilon_{\mu_1, \cdots \mu_s}^* (\hat k, \lambda)  \epsilon^{\mu_1, \cdots \mu_s} (\hat k, \lambda') = \delta_{\lambda \lambda'} \, .
\end{eqnarray}
We will consider the massive case at first. This is especially interesting because (\emph{a}) it is also relevant for bound state instabilitites, to be discussed in Sec. \ref{bound states}, (\emph{b}) it is more difficult to study with traditional methods, because the wave equations do not factorize on a Kerr background (see however~\cite{Rosa:2011my,Pani:2012bp} for recent progress in this direction), and (\emph{c}) in our approach, it is actually algebraically simpler than the massless case due to the lack of gauge invariance, which allows a simpler leading coupling of the field to the dissipative operators. In fact, the leading interaction Hamiltonian that we will consider is
\begin{eqnarray} \la{interaction higher integer spin} 
H_{\rm int} = \alpha \, \Phi_{I_1, \cdots I_s} R^{I_1}{}_{J_1} \cdots R^{I_s}{}_{J_s} \mathcal{O}^{J_1 \cdots J_s} 
\end{eqnarray}
which is clearly not gauge invariant because the operator $\mathcal{O}^{J_1 \cdots J_s}$ is not related to any conserved quantity. Unlike in the simpler scalar case, here we have also included a coupling $\alpha$ that can in principle depend on the mass $\mu$ and the spin $s$ of the field. This coupling is necessary to ensure that our results admit a smooth massless limit. 
In particular, $\alpha$ should vanish when $\mu = 0$ because in this limit the action must become gauge invariant. 

In order to illustrate this point, let us digress for a moment and determine how the coefficient $\alpha$ depends on $\mu$ in the particular case of a spin-1 particle. To this end, we will use the fact that at energies much larger than the mass $\mu$ of the spin-1 particle (but small enough that the spinning object can still be treated as point-like) the longitudinal component must behave like the Goldstone mode of a spontaneously broken $U(1)$ symmetry~\cite{Cornwall:1974km,Vayonakis:1976vz}.\footnote{See also Sec. 21.2 of~\cite{Peskin:1995ev} for a more modern discussion of this equivalence.}

In order to state this equivalence more precisely, let's consider the action for a free massive spin-1 particle, and formally restore gauge invariance by performing the St\"uckelberg replacement 
\be \label{Stueckelberg}
\Phi_\nu \to \Phi_\nu  +\frac{\partial_\nu \phi}{\mu} \, .
\ee
This should \emph{not} be interpreted as a decomposition of $\Phi_\nu$ into its transverse and longitudinal components, but rather as a procedure that yields a new Lagrangian~\cite{Hinterbichler:2011tt}, namely
\be \la{S Stueckelberg}
S= \int d^4 x \l\{ -\frac{1}{4}F_{\mu\nu} F^{\mu\nu} -\frac{\mu^2}{2}\l(\Phi_\nu +\frac{\partial_\nu \phi}{\mu}\r)\l(\Phi^\nu +\frac{\partial^\nu \phi }{\mu}\r)  \r\} \, ,
\ee
that is invariant under the gauge transformation $\Phi_\nu \to \Phi_\nu - \d_\nu \varepsilon, \phi \to \phi + \mu\, \varepsilon$. It is easy to see that one can fix the gauge by setting  $\phi=0$, in which case the usual Proca Lagrangian is recovered. The advantage of introducing explicitly an extra degree of freedom via the replacement (\ref{Stueckelberg}) is that now the action (\ref{S Stueckelberg}) admits a smooth $\mu \to 0$ limit in which  the number of degrees of freedom is preserved. In fact, when $\mu = 0$ we obtain the action for a free massless spin-1 field and a free massless spin-0 field, for a total of three degrees of freedom. As advertised, the longitudinal component of the spin 1 field turns into a $U(1)$ Goldstone mode when the mass $\mu$ is negligible.

Let us now consider a coupling between the field $\Phi_\mu$ and an external source $J_\nu$ that is \emph{not} a conserved current, and perform once again the St\"uckelberg replacement (\ref{Stueckelberg}) to obtain
\begin{eqnarray} \la{S int Stuckelberg}
S_{\rm int}  = \alpha \int d^4 x \,  \l(\Phi^\nu +\frac{\partial^\nu \phi}{\mu}\r) J_\nu.
\end{eqnarray}
If we fix the gauge by setting $\phi =0$, this interaction becomes precisely of the form (\ref{interaction higher integer spin}) and corresponds to a particular choice for the source $J_\nu$. Since the latter is not conserved, the coupling with the scalar field $\phi$ doesn't vanish, and in fact diverges in the limit $\mu \to 0$ unless the coupling $\alpha$ also depends on $\mu$. In particular, requiring that the longitudinal component is dissipated exactly like a massless scalar field in the limit $\mu \to 0$ implies that $\alpha = \mu$ (up to an overall constant that can always be reabsorbed in the definition of $J_\nu$). Similar arguments can be used to determine the coupling $\alpha$ for higher spin fields. For instance, for a massive spin 2 field~\cite{Hinterbichler:2011tt} it is easy to show that one must have $\alpha = \mu^2$. In the rest of this section, we will continue our analysis by keeping $\alpha$ arbitrary.

When studying scattering of higher spin particles by a spherically symmetric object, it is convenient to work with spherical helicity single-particle states~\cite{landau1971course}. These states are labeled by their frequency $\omega$, the \emph{total} angular momentum $j$, the angular momentum in the $z$-direction $m$, and the helicity $\lambda$. The properties of the states $ | \omega, j, m, \lambda \rangle$ that are relevant for the calculation that follows have been summarized in Appendix \ref{appd}.  As before, we will start by studying the absorption process:
\begin{eqnarray}
(\omega, j, m, \lambda) + X_i \to X_f \, .
\end{eqnarray}
The probability for such process to occur is
\be 
P_{\rm abs} = \sum_{X_f} \fr{|\langle X_f; 0| S | X_i; \omega, j, m, \lambda \rangle |^2}{\langle \omega, j, m, \lambda | \omega, j, m, \lambda \rangle} \, .
\ee
Following steps very similar to those already discussed in the previous sections, it is easy to show that 
\begin{eqnarray}
\text{numerator} &=& \alpha^2\int dt' dt''\langle \mathcal{O}^{J_1 \cdots J_s} (t')\mathcal{O}^{L_1 \cdots L_s} (t'') \rangle  R^{I_1}{}_{J_1} \cdots R^{I_s}{}_{J_s}  R^{K_1}{}_{L_1} \cdots R^{K_s}{}_{L_s}  \times \la{numerator higher integer spin absorption}  \\
&& \qquad \qquad \qquad \qquad \quad  \times \langle \omega, j, m, \lambda | \Phi_{I_1, \cdots I_s} (t') | 0 \rangle \langle 0 | \Phi_{K_1, \cdots K_s} (t'') | \omega, j, m, \lambda \rangle \, . \nonumber 
\end{eqnarray}
The structure of the Fourier transform of the operators $\mathcal{O}^{J_1 \cdots J_s}$ is shown in equation (\ref{fourier 2-pt functions higher integer spin}). Moreover, using eq. (\ref{creation annihilation expansion arbitrary integer spin}) and the results discussed in Appendix \ref{appd}, we find that 
\begin{eqnarray}
\langle \omega, j, m, \lambda | \Phi_{I_1 \cdots I_s} (t') | 0 \rangle &=& \sum_{\lambda'} \int \fr{d^3 k'}{(2 \pi)^3 2 \omega_{k'}}  \, \epsilon_{I_1 \cdots I_s}^* (\hat k', \lambda') \langle \omega, j, m, \lambda |  \mbf k', \lambda' \rangle e^{i \omega_{k'} t'} \nonumber \\
&=& \fr{k \, e^{i \omega t'}}{2 \pi \sqrt{2 \omega v}} \l[ \int d\Omega' \,  _{-\lambda}Y_{jm} (\hat k') \epsilon_{I_1 \cdots I_s} (\hat k', \lambda) \r]^*  \, .
\end{eqnarray}
By exploiting rotational symmetry, it is possible to show that 
\begin{eqnarray}
 \int d\Omega' \,  _{-\lambda}Y_{jm} (\hat k') \epsilon_{I_1 \cdots I_s} (\hat k', \lambda) = \delta_{js} \, \sqrt{\fr{4 \pi j!}{(2j+1)!!}} \, V^m_{I_1 \cdots I_s}, \qquad \qquad (|\lambda| \leqslant s)
\end{eqnarray}
where the quantities $V^m_{I_1, \cdots I_s}$ were first introduced in Sec. \ref{higher multipoles}. Thus, convolution with a spin-weighted spherical harmonic $ _{-\lambda}Y_{jm}$ takes a polarization tensor with spin $j$, helicity $\lambda$ and momentum in the direction of $\hat k$ into one with helicity $m$ and momentum in the $z$-direction (up to an overall normalization). By plugging the last two results in the expression for the numerator (\ref{numerator higher integer spin absorption}) and using the result in eq.  (\ref{property of V}), we eventually find that the absorption probability due to the interaction (\ref{interaction higher integer spin}) is
\begin{eqnarray} \la{absorption probability higher spin}
P_{\rm abs} = \fr{\delta_{js} k^2 \alpha^2 }{(2 \pi)^2 2 \omega v} \l[\fr{4 \pi j!}{(2j+1)!!} \r] \Delta_s (\omega - m \Omega) \, .
\end{eqnarray}

Note that at this order our result does not depend on the helicity $\lambda$, but only on the angular momentum numbers $j$ and $m$. By now, the reader should be familiar with the rest of the argument: the emission probability for a spin $s$ particle is
\begin{eqnarray}
P_{\rm em} = \fr{k^2  \alpha^2 \delta_{js}}{(2 \pi)^2 2 \omega v} \l[\fr{4 \pi s!}{(2s+1)!!} \r]  \Delta_s (m \Omega-\omega) \, ,
\end{eqnarray}
and the relative difference between outgoing and incoming flux is 
\begin{eqnarray} \la{flux change l}
\fr{\Phi_{\rm out}- \Phi_{\rm in}}{\Phi_{\rm in}} = P_{\rm em} - P_{\rm abs} = -   \fr{k^2  \alpha^2 s! \delta_{js}}{2 \pi (2s+1)!! \omega v} \rho_s (m \Omega-\omega) \, .
\end{eqnarray}

In the strictly massless limit, the coefficient $\alpha (\mu, s)$ must vanish because the interaction (\ref{interaction higher integer spin}) is forbidden by gauge invariance. The leading interaction involves the field strength tensors for the higher spin $s$ fields $\Phi_{\mu_1 \cdots \mu_s}$, which are gauge invariant quantities~\cite{Goldberger:2005cd,Porto:2007qi,Endlich:2015mke}. For massless spin one particles (i.e. electromagnetism), the field strength is as usual $F_{\mu \nu} = \d_\mu \Phi_\nu - \d_\nu \Phi_\mu$. Its components have mixed parity, as can be seen directly at the level of the photon degrees of freedom \cite{Thorne:1980ru}.  It can be decomposed into its definite parity states by separating it into electric and magnetic components. The leading couplings (in the rest frame of the dissipative object) are thus given by
\be \label{gauge vector couplings}
H_{int}=E^I R_{I}{}^{J} \mathcal{O}^E_J + B^I R_{I}{}^{J} \mathcal{O}^B_J \, ,
\ee
where $E^I=F^{0I}$ and $B^I=\frac{1}{2} \epsilon^{IJK}F_{JK}$. Notice that the operators $\mathcal{O}^E_J$ and $\mathcal{O}^B_J$ have opposite parity, and consequently their mixed two point function vanishes.

For massless spin two particles (i.e. gravity), the field strength is give by the Weyl tensor $C_{\alpha \beta \mu \nu}$.\footnote{The Weyl tensor is the relevant field strength because couplings involving the traces of the Riemann tensor $R_{\alpha \beta \mu \nu}$ can always be removed by field redefinitions of the metric using the leading order equations of motion $R_{\mu\nu}=0$.} Here too a similar decomposition into states of definite parity can be performed. Upon separating the Weyl tensor into its electric and magnetic parity components the leading coupling (again in the rest frame of the dissipative object) is given by
\be \la{massless graviton couplings}
H_{\rm int}=E^{IJ} R_{I}{}^{N}R_{J}{}^{M} \mathcal{O}^E_{NM} + B^{IJ} R_{I}{}^{N}R_{J}{}^{M} \mathcal{O}^B_{NM}
\ee
where $E_{IJ}=C_{0I0J}$ and $B_{IJ}=\frac{1}{2}\epsilon_{0INM}C_{0J}{}^{NM}$ \cite{Endlich:2015mke}. Just as in the spin 1 case, the two operators $\mathcal{O}^E_{NM}$ and $ \mathcal{O}^B_{NM}$ do not mix with each other at the level of the two point function. Note moreover that the couplings (\ref{massless graviton couplings}) describing superradiant scattering of long wavelength gravitons are also responsible for tidal dissipation~\cite{Endlich:2015mke}. This makes precise the conjectured ``correspondence'' between tidal dissipation and superradiant effects discussed in~\cite{Glampedakis:2013jya} and references therein.

As we can see from the above couplings, the leading interactions for massless spin $s$ particles are suppressed by $s$ additional powers of $\omega$ compared to the one in eq. (\ref{interaction higher integer spin}). Therefore, the resulting absorption and emission probabilities pick up an additional factor of $\omega^{2s}$ at low frequencies compared to the massive case. Explicit results for the absorption of massless particles with integer spin by a Schwarzschild black hole confirm this expectation~\cite{Page:1976df}.


\section{Superradiant instability of bound states} \la{bound states}

With a few minor modifications, the same techniques developed in the previous sections to study superradiant scattering can be used to discuss the superradiant instability of bound states around spinning objects. Our approach is valid whenever the Compton wavelength of the particle forming the bound state is much larger than the size of the spinning object, in which case the latter can be treated as point-like. 

\subsection{Scalar bound states} \la{sec: scalar bound state}

 For simplicity, we will first consider the case of a spin 0 particle with non-zero\footnote{Note that we can have bound states only if the mass of the scalar field is non-zero.} mass $\mu$. In the presence of bound states, the  expansion of a scalar field in terms of creation and annihilation operators takes the form:
\begin{eqnarray} \la{phi expansion bound states}
\hat \phi = \sum_{n \ell m} \fr{1}{\sqrt{2 E_{n\ell m}}} \bigg\{ \hat a_{n\ell m} \, f_{n\ell m} (r, \theta, \varphi) e^{- i E_{n\ell m} t} + \hat a_{n\ell m}^\dag \, f_{n\ell m}^* (r, \theta, \varphi) e^{i E_{n\ell m} t}   \bigg\} + \cdots  \, ,
\end{eqnarray}
where the quantum numbers $n, \ell, m$ label the different bound states, and the dots stand for the usual sum over creation and annihilation operators of asymptotic states with definite momentum. It will be convenient to choose the normalization of the bound states so that they are orthonormal. The factor of $(2 E_{n\ell m})^{-1/2}$ in eq. (\ref{phi expansion bound states}) was judiciously inserted in such a way that
\begin{eqnarray} \la{normalization of f}
 \displaystyle\int d \Omega \, dr \, r^2  f_{n\ell m} (r, \theta, \varphi) f_{n'\ell'm'}^* (r, \theta, \varphi) =  \delta_{nn'} \delta_{\ell\ell'} \delta_{mm'}  \, . 
\end{eqnarray}

Clearly, in order to find non-trivial bound state solutions we must take interactions into account. At the same time, at distances much larger than the size of the spinning object the leading contribution comes from the $1/r$ interaction.\footnote{Higher order corrections can be systematically taken into account within the effective theory~\cite{Goldberger:2004jt,Goldberger:2007hy,Porto:2016pyg}---remember that one of our expansion parameters is $v^2$, which in turn is proportional to $1/ r$ by the virial theorem.} In this limit, we can always factorize the mode functions as $f_{n\ell m} (r, \theta, \varphi) = R_{n\ell} (r) Y_{\ell m} (\theta, \varphi)$, and the functions $\psi_{n\ell m} \equiv r f_{n\ell m}$ satisfy the same 1D Schr\"odinger equation that describes the Hydrogen atom~\cite{Detweiler:1980uk}. In our case, however, the non-relativistic binding energy $E$ replaced by $(E^2_{n\ell m} - \mu^2)/2 \mu$. Therefore, at lowest order in the interaction the energy eigenvalue $E_{n\ell m}$ depends only on $n$~\cite{Sakurai:2011zz}: 
\begin{eqnarray} \la{hydrogen-like spectrum}
E_{n\ell m}^2 \simeq \mu^2 \l[ 1 - \fr{(G M \mu)^2}{n^2} \r], \qquad \qquad \ell + 1 \leqslant n \, .
\end{eqnarray}
In this equation, we have used the fact that the strength of the gravitational interaction responsible for the bound states is given by $G M \mu$. In the regime we are interested in (where the Compton wavelength of the scalar field is much larger than the size of the compact object) we have that $G M \mu \ll1$, and therefore we can safely neglect the binding energy and set $E_{n\ell m} \simeq \mu$.

We will be interested in calculating the absorption and emission probability of a bound state $| n, \ell, m \rangle \equiv \hat a_{n\ell m}^\dag | 0 \rangle$ by the spinning object.  Let's first consider the probability for the absorption process
\begin{eqnarray}
X_i + (n, \ell, m) \to X_f \, .
\end{eqnarray}
If we don't care about the final state $X_f$ of the spinning object the probability is equal~to:
\begin{eqnarray}
P_{\rm abs} = \sum_{X_f} | \langle X_f ; 0 | S | X_i ; n, \ell, m \rangle |^2 \, ,
\end{eqnarray}
with the $S$-matrix defined in equation (\ref{S}). Once again, we will restrict our attention to the interaction Hamiltonian (\ref{Hint spin 0}), since this will yield the leading dissipative contribution for modes with $\ell = 1$, which in turn are the most unstable ones~\cite{Detweiler:1980uk}. In this case, steps analogous to those followed in section \ref{spin_0} yield the following expression for $P_{\rm abs}$:
\begin{eqnarray}
P_{\rm abs} =  \int \fr{d \omega}{2 \pi}\fr{ \Delta (\omega)}{2 \mu} \l| \int dt \, \d^I f_{n\ell m} (r=0) R_I{}^J(t)e^{i(\omega - \mu)t}\r|^2 \, ,
\end{eqnarray}
where we have used the fact that $E_{nlm} \simeq \mu$. To proceed further, we can use the identity~\cite{Thorne:1980ru}
\be \la{Solomon's trick}
Y_{\ell m}(\theta, \vphi)=\sqrt{\fr{(2\ell+1)!!}{4 \pi \ell!}}V^m_{I_1 \cdots  I_\ell} \hat r^{I_1} \cdots \hat r^{I_\ell} \, 
\ee
to rewrite the $\d^I f_{n\ell m} (r=0)$ as follows:
\begin{eqnarray}
\d^I f_{n\ell m} (r=0) = \fr{1}{\sqrt{4 \pi}} \l\{ \delta_\ell^0 \delta_m^0 \, \hat r^I  \l.\d_r R_{n0} \r|_{r=0} + \sqrt{3} \, \delta_\ell^1 V_m^I \l. \fr{R_{n1}}{r} \r|_{r=0}  \r\}.
\end{eqnarray}
Hence, we see that the the interaction (\ref{Hint spin 0}) affects not only modes with $\ell=1$, but also with $\ell =0$. This should be contrasted with the dissipation of asymptotic states discussed in the previous sections, in which case the interaction (\ref{Hint spin 0}) affected exclusively modes with orbital angular momentum $\ell =1$. Notice however that this interaction still provides the dominant contribution for $\ell =1$ modes. The $\ell =0$  contribution is instead subleading compared to the one coming from the interaction with no derivatives $H_{\rm int} = \int dt \, \phi \, \mathcal O$.\footnote{It is in fact easy to check that this coupling would give an absorption rate for $\ell = 0$ modes that scales like $\Gamma \propto \Delta (\mu) (G M \mu^2)^{3} / \mu$. The coupling (\ref{Hint spin 0}) that we are considering gives a correction to this result which is of the same order as the rate for $\ell =1$ modes in eq. (\ref{gamma abs bound scalar}) with $\Omega = 0$.} More in general, interactions with $\ell$ derivatives will yield the leading dissipative contribution for modes with orbital angular momentum $\ell$, and subleading dissipative corrections for modes with angular momentum $< \ell$. It is therefore still correct to consider interactions with $\ell$ derivatives to extract the leading behavior of modes with orbital angular momentum $\ell$.

Since $\ell=0$ modes are not susceptible to superradiance, from now on we will set $\ell =1$. Moreover, the overall amplitude of the functions $R_{n\ell}(r)$ decreases with increasing $n$ once these functions are normalized according to eq. (\ref{normalization of f})~\cite{Sakurai:2011zz}. Hence, the largest absorption rate occurs for the mode with $n=2$ (the same goes for the emission rate, which we will turn to in a moment), and therefore we will focus on such a mode for the remaining of this section. Using the fact that for small values of $r$~\cite{Sakurai:2011zz} 
\begin{eqnarray}
\label{R_21}
R_{21}(r) \simeq \l( \fr{GM \mu^2}{2} \r)^{5/2} \fr{2 r}{\sqrt{3}},
\end{eqnarray}
we can easily calculate $\d^I f_{21m}$ to find
\begin{eqnarray} \la{gamma abs bound scalar}
\Gamma_{\rm abs} = \fr{1}{2 \pi \mu} \l( \fr{GM \mu^2}{2} \r)^{5} \Delta (\mu - m \Omega ), \qquad\qquad (n=2, \,\,\ell=1) \, .
\end{eqnarray}
A completely analogous calculation shows that the spontaneous emission rate is instead
\begin{eqnarray}
\Gamma_{\rm em} = \fr{P_{\rm em}}{T} \simeq \fr{1}{2 \pi \mu} \l( \fr{GM \mu^2}{2} \r)^{5} \Delta (m \Omega - \mu), \qquad\qquad (n=2, \,\, \ell=1) \, .
\end{eqnarray}

Whether or not the spinning object is ultimately unstable due to accretion of the bound state under consideration depends on the relative magnitude of the absorption and emission rates. The difference between these two rates for the $n=2, \ell=1$ mode is
\begin{eqnarray}
\Delta \Gamma = \Gamma_{\rm em} - \Gamma_{\rm abs} \simeq  \l( \fr{GM \mu^2}{2} \r)^{5} \fr{ \rho (m \Omega - \mu) }{2 \pi \mu}\simeq  \l( \fr{GM \mu^2}{2} \r)^{5} \fr{ ( m \Omega - \mu) \gamma }{2 \pi \mu} \, ,
\end{eqnarray}
where in the last step we have used the low frequency limit of the spectral density $\rho$. A positive value of $\Delta \Gamma$ signals an instability, because the rate of production of particles in a bound state is larger than the rate of their absorption. Such instability occurs for superradiant modes, i.e. for bound states such that $\mu - m \Omega <0$, which can only be achieved if $m=1$ and $\Omega > \mu$.

In the case of a black hole, we have previously found that $\gamma = \frac{2}{3} \pi r_s^4$ by matching the probability for absorption of an incoming spherical wave by a Schwarzschild black hole. Using this value, we can find immediately  the instability rate for a Kerr black hole due to the production of bound states with spin 0 particles with mass $\mu \ll \Omega$:
\begin{eqnarray} \la{scalar bound state instability rate}
\Delta \Gamma \simeq \fr{ \l(G M \mu\r)^9}{6} \Omega  \, .
\end{eqnarray}
This result agrees perfectly with the explicit analytic calculation based on the Kerr metric~\cite{Detweiler:1980uk}.\footnote{See also~\cite{Ternov:1978gq} for a calculation of the absorption rate of bound states by Schwarzschild and Kerr black holes.} Notice however that in our approach we never had to solve a wave equation on a Kerr background. Knowing the form of the mode functions and the dissipative coefficient $\gamma$ for the non-spinning object was enough to find the instability rate for the spinning object. This was possible because, by taking the point-like limit and modeling dissipation as interactions with composite operators, we managed to disentangle the problem of calculating the instability rate from that of finding the mode functions. Moreover, the latter task becomes especially easy in the Newtonian limit, which is always a good approximation for Compton wavelengths much larger than the size of the spinning object.  Because gravitational interactions are spin-independent in the Newtonian limit, our approach can be extended also to higher integer spin particles, for which the usual approach is ill-suited because the wave equation is not factorizable on a Kerr background (see however~\cite{Pani:2012bp} for a recent approach based on perturbation theory). As an illustration, we will now consider the particularly interesting spin 1 case.

\subsection{Vector bound states} \la{vector bound states}

In the presence of a Newtonian potential, helicity is no longer a good quantum number because translations are spontaneously broken by the center of the potental. Therefore, for massive vector fields we will label the bound states using instead the \emph{orbital} angular momentum~$\ell$, together with the quantum numbers $n, j, m$. According to the usual rules of addition of angular momenta, the only allowed values of $\ell$ for a given value of $j$ are $\ell= j -1, j, j +1$. Introducing the ``collective'' index $\beta \equiv (n, j, m, \ell)$, we can expand a vector field in terms of creation and annihilation operators as follows: 
\begin{eqnarray} \la{phi mu expansion bound states}
\hat \Phi_{\mu}  = \sum_{\beta} \fr{1}{\sqrt{2 E_{\beta}}} \bigg\{ \hat a_{\beta} \, f^{\beta}_{\mu} (r, \theta, \varphi) e^{- i E_{\beta} t} + \hat a_{\beta}^\dag \, f^{\beta *}_{\mu} (r, \theta, \varphi) e^{i E_{\beta} t}   \bigg\} + \cdots  \, ,
\end{eqnarray}
where the dots stand for the usual continuum states. The vector field $\Phi_\mu$ satisfies the Proca equation, which is also equivalent to the set of equations
\be  \la{proca equivalent}
\nabla_\mu \Phi^\mu = 0, \qquad \qquad\qquad  \square \Phi^\mu = \mu^2 \Phi^\mu \, .
\ee
The first equation in particular is a constraint ensuring that there are only three physical polarizations. These equations are valid on an arbitrary gravitational background, but in the limit of small $\mu$ (again, compared to the inverse size of the spinning object) we can work in the Newtonian limit, as we did in the spin 0 case.

It is useful to decompose the mode functions for the bound states of $\Phi_\mu$ that appear in eq. (\ref{phi mu expansion bound states}) as follows:
\be \la{mode functions of spin 1 bound state}
f_\mu^{n j m \ell} (r, \theta, \vphi)= \bigg( i S_{n j \ell} (r) Y_{jm} (\theta, \vphi),  R_{n\ell}(r)\mbf{Y}^{\ell,jm} (\theta, \vphi) \bigg) \, .
\ee
where the $ \mbf{Y}^{\ell,jm}$'s are vector spherical harmonics~\cite{Thorne:1980ru}. They are defined by combining the spin 1 polarization vectors defined in (\ref{def V}) with ordinary spherical harmonics with orbital angular momentum $\ell$ to create quantities with quantum numbers $j, m$:
\be
\label{vec_spherical_def}
\mbf{Y}^{\ell,jm} (\theta, \vphi) =  \sum_{m'=-1}^1 \sum_{m''=-\ell}^\ell C_{1 \ell}(j, m; m', m'') \mbf{V}^{m'} Y_{\ell m''}(\theta, \vphi) \, ,
\ee
The coefficients of this linear combination are the usual Clebsh-Gordan coefficients~\cite{Sakurai:2011zz}. 

The functions $S_{nj \ell} (r)$ in eq. (\ref{mode functions of spin 1 bound state}) can be expressed in terms of the $R_{n\ell}(r)$'s by solving the constraint in eq. (\ref{proca equivalent}) (see Appendix \ref{higher l bound states} for an explicit expressions). Then, as the vector spherical harmonics are eigenfunctions of the orbital angular momentum operator $\hat L^2\equiv - [ \fr{1}{\sin^2 \theta} \d_\vphi^2 + \sin \theta \d_\theta (\sin\theta \d_\theta)]$, the remaining equations reduce in the Newtonian limit to a single Hydrogen-like Schr\"odinger equation for the functions $\psi_{n\ell} (r) \equiv r R_{n\ell}(r)$. In particular, the energy eigenvalues $E_\beta$ are still of the form given in eq. (\ref{hydrogen-like spectrum}) and therefore depend only on the quantum number $n$.

The dominant dissipative interaction at low-energies is determined by the values of $j$ and $\ell$. From this viewpoint, the vector case is more involved than the scalar one, where the value of the orbital angular momentum $\ell$ was sufficient to determine the relevant interaction. In what follows we will discuss only the modes with $j =1$. We refer the reader interested in higher values of $j$ to Appendix \ref{higher l bound states}. However, the $j=1,\ell=0$ mode turns out to have the largest instability rate, and thus it singlehandedly determines (parametrically, at least) the time scale for superradiant instability due to vector bound states.

Modes with $j=1$ are dissipated by composite operators that carry a single index. In particular, the three interactions that we will need to consider when $j=1$ have already been introduced in Sec. \ref{higher spins}, and are\footnote{There is one more interaction that one could write down that involves an operators with a single index, namely $\mu \, \d_i \Phi^0 R^I{}_J \tilde{\mathcal{O}}^J$. The factor of $\mu$ is again necessary to ensure a smooth $\mu \to 0$ limit. However, it is easy to see that this interaction would be further suppressed compared to the ones in eq. (\ref{3 interactions}), because it is equal to $\mu \, (\d_0 \Phi_I - E_I ) R^I{}_J \tilde{\mathcal{O}}^J$. } 
\begin{eqnarray} \la{3 interactions}
H_{\rm int }  = \mu \, \Phi^I R_{I}{}^{J} \mathcal{O}_J, \qquad \qquad H_{\rm int } = E^I R_{I}{}^{J} \mathcal{O}^E_J, \qquad \qquad H_{\rm int }  = B^I R_{I}{}^{J} \mathcal{O}^B_J \, ,
\end{eqnarray}
where $E^I=F^{0I}$, $B^I=\frac{1}{2} \epsilon^{IJK}F_{JK}$, and $F_{\mu\nu}$ is the field strength of $\Phi_\mu$. The three operators $\mathcal{O}_J, \mathcal{O}^E_J$ and $\mathcal{O}^B_J$ have different transformation properties under parity and time reversal, as shown in Table \ref{tab:1}. As such they do not mix at the level of the two-point function, which is all we will need for our purposes. 

\begin{table}[t]
\begin{center}
\begin{tabular}{c|c|c|c}
& $\mathcal{O}_J$ & $ \mathcal{O}^E_J$ &  $\mathcal{O}^B_J$ \\
\hline
$P$ & $-$ & $-$ &  $+$ \\
\hline
$T$ & $-$ & $+$ & $-$ \\
\end{tabular}
\end{center}
\caption{Parity and time reversal properties of $j=1$ composite operators.}
\label{tab:1}
\end{table}%

While $j$ is associated with the number of indices carried by the composite operator, $\ell$ is determined by the number of spatial derivatives acting on $\Phi^I$. Before applying this rule of thumb, we should solve the constraint $\nabla_\mu \Phi^\mu = 0$ schematically to obtain $ \d_I \Phi^I \sim \d^0 \Phi^0 \sim \mu \Phi^0$ and therefore $E^I = \d^0 \Phi^I - \d^I \Phi^0 \sim \mu \Phi^I - \d^I \d_J \Phi^J  / \mu$. We should also remember that $B^I = \epsilon^{IJK} \d_J \Phi_K$. Hence, the $\ell = 2$ and $\ell = 1$ modes will be dissipated exclusively by the second and third interaction in (\ref{3 interactions}) respectively, whereas the $\ell =0$ mode will receive a contribution from each of the first two interactions. Note that parity considerations would also be sufficient to conclude that the operator $\mathcal{O}^B_J$ must be responsible for dissipating the $\ell =1$ mode, since the latter is even under parity~\cite{Thorne:1980ru}. 

By proceeding in complete analogy with the scalar case and using the ``technology'' developed in Secs. \ref{higher spins} and \ref{sec: scalar bound state}, we can easily calculate the instability rates for the three modes with $j=1$ and find
\begin{eqnarray}
\Delta \Gamma (j=1, \ell = 0) &\simeq& (\gamma + \gamma_E)  (GM \mu^2)^3 \fr{\mu \, \Omega}{2 \pi} \, , \\
\Delta \Gamma (j=1, \ell = 1) &\simeq&  \gamma_B \l( \fr{GM \mu^2}{2} \r)^5 \fr{\Omega}{ \pi \mu} \, , \\
\Delta \Gamma (j=1, \ell = 2) &\simeq& \fr{10 }{9} \, \gamma_E \, \l( \fr{GM \mu^2}{3} \r)^7 \fr{\Omega}{ \pi \mu^3} \, .
\end{eqnarray}
The coefficients $\gamma, \gamma_E$ and $\gamma_B$ arise by expanding the spectral densities of $\mathcal{O}_I,\mathcal{O}_I^E$ and $\mathcal{O}_I^B$ respectively, as shown in eq. (\ref{rho small omega}). Crucially, these three coefficients determine not only the instability rate due to vector bound states, but also the absorption probabilities of the three massive vector polarizations in a superradiant scattering process. Moreover, at energies larger than $\mu$ but still small enough that the object can be treated as point-like, these probabilities should reduce to those for absorbing a massless vector and a scalar with $\ell=1$. These probabilities are well known for a black hole~\cite{Page:1976df}, and by matching onto them we find
\begin{eqnarray}
\gamma_E = \gamma_B = 4 \gamma = \fr{8 \pi}{3} \, r_s^4.
\end{eqnarray}
By using these results we can determine \emph{exactly} the instability rates at leading order:
\begin{eqnarray}
\Delta \Gamma (j=1, \ell = 0) &\simeq& \fr{80}{3} (G M \mu)^7 \Omega \, , \la{leading vector instability rate} \\ 
\Delta \Gamma (j=1, \ell = 1) &\simeq& \fr{4}{3}(G M \mu)^9 \Omega \, , \\
\Delta \Gamma (j=1, \ell = 2) &\simeq& \fr{1280}{59049}(G M \mu)^{11} \Omega \, .
\end{eqnarray}

The parametric dependence of these rates on the combination $(G M\mu)$ is consistent with that of the absorption rates calculated in~\cite{Rosa:2011my,Pani:2012bp}. This agreement persists also for higher values of $j$---see Appendix \ref{higher l bound states} for more details---and it confirms the scaling rule proposed in~\cite{Rosa:2011my,Pani:2012bp}:\footnote{In comparing our results with the ones in~\cite{Rosa:2011my, Pani:2012bp} for arbitrary values of $\ell$ and $j$, one should be aware that $\ell_{\rm there} = j_{\rm here}$ and $j_{\rm there} = \ell_{\rm here}$. We find our notation to be more in line with the traditional one, since indeed our $j$ and $\ell$ are respectively the total and orbital angular momenta.\la{footnote notation}}
\begin{eqnarray}
\Delta \Gamma \propto (G M\mu)^{5 + 2 j + 2 \ell}.
\end{eqnarray}
This scaling was originally derived using mostly numerical methods, which in the limit $G M \mu \ll1$ become increasingly unreliable~\cite{Pani:2012bp}.\footnote{On the other hand, numerical methods are crucial to explore the regime $G M \mu \gtrsim 1$ in which our perturbative approach ceases to be applicable.} What is remarkable is that our EFT approach yields not only the parametric scaling, but also the exact overall coefficient. As anticipated earlier, the most unstable mode is the one with $j =1$ and $\ell =0$, and its instability rate (\ref{leading vector instability rate}) is parametrically larger than the one for scalar bound states in eq. (\ref{scalar bound state instability rate}). To conclude, we should also mention that the procedure carried out in this section for vector bound states could be easily repeated for massive higher spin particles using tensor spherical harmonics~\cite{Winter1982}.


\section{Conclusions} \la{conclusions}

In this work, we have discussed a modern, perturbative approach to (rotational) superradiance based on effective field theory techniques. Our formalism describes slowly spinning objects interacting with particles of any mass and spin whose energy is much smaller than the inverse size of the object (in natural units). Within this framework, we show unambiguously that superradiance is not peculiar to the Kerr solution, but rather is a generic feature of any dissipative rotating object. As such, our results apply also to astrophysical systems other than black holes, which generically are not described by a Kerr metric.  For simplicity, we have restricted our attention to spherically symmetric objects, although it would be interesting and in principle straightforward to relax this assumption.

We argued that, at lowest order in perturbation theory, the \emph{same} parameters determine (1) the absorption probability of a particle with a given spin, mass and polarization (2) the superradiant amplification rate of a beam of such particles, (3) the rate of superradiant instability due to formation of bound states, and (4) the relaxation time scale due to vacuum friction. These parameters can be extracted from analytic (for black holes) or numeric (for other astrophysical objects) calculations of the absorption probability in the (relatively simpler) static limit, and then used in the more complicated spinning case. This is an improvement on a similar EFT treatment given in~\cite{Porto:2007qi}, which required an additional matching procedure for the spinning case. For spin 0 particles, the same parameters describe both the massless and massive case. This is not the case for higher spin particles, which require two distinct sets of parameters.

Within our framework, we were able to unify a variety of results that were previously scattered across the literature, as well as to derive some interesting new ones. In particular, we calculated the absorption probability and superradiant amplification rate for massive particles with \emph{arbitrary} integer spin scattering off a \emph{generic} spinning object (that is, not necessarily a black hole).

We also calculated the instability rate due to formation of bound states with massive spin 0 and spin 1 particles. By working at lowest order in the gravitational coupling, we showed that the former scales like $\l(G M \mu\r)^9$ and the latter like $(G M \mu)^7$ for small values of $G M \mu$. This estimate was obtained by studying states with $\ell =1$ and $\ell =0$ respectively, as these are the most unstable ones. There is however no conceptual obstacle to extending our calculations to higher values of $\ell$. Our results for vector bound states are consistent with the recent numerical results of~\cite{Rosa:2011my,Pani:2012bp}, but disagree with earlier results obtained analytically in~\cite{Galtsov:1984nb}.

Finally, when combined with our previous work~\cite{Endlich:2015mke}, our results make explicit the ``correspondence'' between tidal distortion and gravitational superradiance put forward in~\cite{Glampedakis:2013jya}. From our EFT perspective, the connection between these two seemingly unrelated phenomena follows immediately from the fact that they are governed by the same dissipative couplings in the effective action.


\section*{Acknowledgments}

We would like to thank Asimina Arvanitaki, Masha Baryakhtar, Emanuele Berti, Chris Brust, Vitor Cardoso, Luca Delacr\'etaz, Savas Dimopoulos, Sam Dolan, Sergei Dubovsky, Walter Goldberger, Kurt Hinterbichler, Junwu Huang, Robert Lasenby, Mehrdad Mirbabayi, Alberto Nicolis, Paolo Pani, Jo\~ao Rosa, Rachel Rosen, Ira Rothstein and Mae Teo for interesting and helpful conversations. We also acknowledge the Abdus Salam International Centre for Theoretical Physics, the Perimeter Institute, and the Sitka Sound Science Center for their generous hospitality during different stages of this collaboration. The work of R.P. was supported by the US Department of Energy under contract DE-FG02-11ER41743 and by the (HEP) Award DE-SC0013528.


\appendix

\section{Normalization of single particle states} \la{appa}

We are using the relativistic normalization for 1-particle states that are eigenstates of momentum, i.e.
\be
\langle \mbf{q} | \mbf{k} \rangle = 2 \omega_k (2 \pi)^3 \delta (\mbf{q} - \mbf{k}) \, .
\ee
Since the one particle states can also be defined as $| \mbf{k} \rangle = a^\dag_\mbf{k} |0 \rangle$, this normalization corresponds to having the following algebra of creation and annihilation:
\be
[ a_{\mbf{k}},a^\dag_{\mbf{q}} ] = (2 \pi)^3 \, 2 \omega_{k} \, \delta (\mbf{k}-\mbf{q}) \, .
\ee
Moreover, with this normalization the completeness relation reads
\be
\int \fr{d^3 k}{(2 \pi)^3 2 \omega_k} | \mbf{k} \rangle \langle \mbf{k} | = \mbf{1} \, .
\ee
For a massless particle, we have of course $\omega_k = k$, but in what follows we will try to keep the equations general for as long as possible. We are now interested in introducing a different basis for 1-particle states, namely $| \omega, \ell, m \rangle$. Let's now discuss the normalization of these states. First, notice that 
\be
(\omega_k - \omega)  \langle \mbf{k} |  \omega, \ell, m \rangle = \langle \mbf{k} | (H - \omega) | \omega, \ell, m \rangle = 0 \, .
\ee
Thus, we we can assume that 
\be
\langle \mbf{k} |  \omega, \ell, m \rangle = N_{\omega, \ell,m} \delta(\omega_k - \omega) Y_\ell^m (\hat{\mbf{k}}) \, .
\ee
We will assume that the spherical harmonics $ Y_\ell^m (\hat{\mbf{k}})$ are normalized in such a way that 
\be
\int d \Omega \,Y_{\ell'}^{m'} (\hat{\mbf{k}})^* Y_\ell^m (\hat{\mbf{k}}) = \delta_{\ell\ell'} \delta^{m m'} \, .
\ee
Then, we have
\ba
\langle \omega', \ell', m' |  \omega, \ell, m \rangle &=& \int \fr{d^3 k}{(2 \pi)^3 2 \omega_k} \langle \omega', \ell', m'  | \mbf{k} \rangle \langle \mbf{k} | \omega, \ell, m \rangle \nonumber \\
&=& \int \fr{d \Omega \, d k \, k^2}{(2 \pi)^3 2 \omega_k}  \delta(\omega_k - \omega)   \delta(\omega_k - \omega')  Y_{\ell'}^{m'} (\hat{\mbf{k}})^* Y_\ell^m (\hat{\mbf{k}}) N_{\omega, \ell,m} N^*_{\omega', \ell',m'} \nonumber \\
&=& \delta (\omega - \omega') \delta_{\ell\ell'} \delta^{m m'} \int \fr{d \omega_k \, k^2}{(2 \pi)^3 2 \omega_k v }  \delta(\omega_k - \omega) |N_{\omega, \ell,m}|^2 \nonumber \\
&=& \delta (\omega - \omega') \delta_{\ell\ell'} \delta^{m m'} |N_{\omega, \ell,m}|^2 \fr{k^2}{2 \omega v (2 \pi)^3}
\ea
where we have introduced the group velocity $v = d\omega / dk$, while $M$ is the mass of the particle. For simplicity, we will choose $N_{\omega, \ell,m}$ in such a way that
\be
|N_{\omega, \ell,m}|^2 \fr{k^2}{2 \omega v (2 \pi)^3} \equiv 2 \pi  \, .
\ee
With this convention, we have 
\be \la{normalization of spherical states}
\langle \omega', \ell', m' |  \omega, \ell, m \rangle = 2 \pi \delta (\omega - \omega') \delta_{\ell\ell'} \delta^{m m'}
\ee
and 
\be \la{k to omega l m }
\langle \mbf{k} |  \omega, \ell, m \rangle =  \fr{\sqrt{2 \omega v}}{k}\,  (2 \pi)^2 \delta(\omega_k - \omega) Y_\ell^m (\hat{\mbf{k}}) \, .
\ee
This latter equation means that 
\be
|  \omega, \ell, m \rangle = \int \fr{d^3 k}{(2 \pi)^3 2 \omega_k} | \mbf{k} \rangle \langle \mbf{k} |  \omega, \ell, m \rangle = \fr{k}{\sqrt{2 \omega v}}\,  \int  \fr{d \Omega}{2 \pi} \, Y_\ell^m (\hat{\mbf{k}})  | \omega, \hat{\mbf{k}} \rangle \, ,
\ee
where once again carried out the integral over $k$ by turning it into an integral over $\omega_k$.
Consequently, we can write the creation operator for a spherical wave single particle state of frequency $\omega_k$ and angular momentum quantum numbers $\ell$ and $m$ as simply
\be
a^\dagger_{\omega_k \ell m} =  \fr{k}{\sqrt{2\omega_k v}} \int \fr{d \Omega}{2 \pi} \, Y_\ell^m (\hat{\mbf{k}}) \,  a^\dagger_{ \mbf{k}}  \, .
\ee
Similar manipulations yield also
\be
a^\dagger_{ \mbf{k}}=\sum_{\ell m}2\pi \fr{\sqrt{2 \omega_k v }}{k}Y_\ell^m (\hat{\mbf{k}})^* \, a^\dagger_{\omega_k \ell m} \, .
\ee
One can then check directly that the commutation relations in the spherical wave basis are simply
\be
\left[a_{\omega \ell m} \, , \, a^\dagger_{\omega' \ell' m'} \right] = 2 \pi \delta (\omega - \omega') \delta_{\ell\ell'} \delta^{m m'}
\ee
and consistent with our normalization of states when we define $|  \omega, \ell, m \rangle \equiv a^\dagger_{\omega \ell m}| 0 \rangle$.
Finally, let us derive the form of the completeness relation with our normalization:
\ba
\mbf{1} &\equiv& \sum_{\ell, m} \int_M^{\infty} \fr{d \omega}{2 \pi} M_{\omega, \ell, m}  | \omega, \ell, m \rangle \langle \omega, \ell, m |  \nonumber \\ 
&=&  \sum_{\ell, m} \int_M^{\infty}  \fr{d \omega}{2 \pi} M_{\omega, \ell, m} \fr{k^2}{2 \omega v} \int \fr{d \Omega}{2 \pi} \fr{d \Omega'}{2 \pi} Y_\ell^m (\hat{\mbf{k}}) Y_\ell^m (\hat{\mbf{k}'})^* | \omega, \hat{\mbf{k}} \rangle \langle \omega, \hat{\mbf{k}}'| \nonumber \\
&=& \int_M^{\infty}  \fr{d \Omega d \omega_k \, k^2}{(2 \pi)^3 2 \omega_k v} M_{\omega_k, \ell, m}  | \omega_k, \hat{\mbf{k}} \rangle \langle \omega_k, \hat{\mbf{k}}| \qquad \Longrightarrow \qquad M_{\omega, \ell, m} = 1 \, .
\ea
Thus, the completeness relation in the $| \omega, \ell, m \rangle$ basis is
\be
\sum_{\ell, m} \int_M^{\infty}  \fr{d \omega}{2 \pi} \, | \omega, \ell, m \rangle \langle \omega, \ell, m |  = \mbf{1} \, .
\ee

\section{Cross section and decay rate} \la{appb}

For completeness, we will re-derive the relation (\ref{cross section l=1}) between the cross section and the absorption probability (\ref{final result absorption}) for a particle with an arbitrary dispersion relation. To this end, it is convenient to first consider a different process, namely 
\be
X_i + \mbf{k} \to X_f \, .
\ee
The probability $P'$ for the inclusive process (where we sum over all possible final configurations $X_f$) is related to the cross section $\sigma$ by
\be \la{sigmak}
\sigma(\mbf{k}) = \fr{P'}{T \times \text{flux}}, \qquad\qquad  \text{flux} = \fr{v}{V} \, ,
\ee
where $v$ is the group velocity of the particle. Notice also that we are showing explicitly the fact that the cross section depends on the incoming momentum $\mbf{k}$. Now, keeping in mind how the normalization of states and the definition of $P'$, we find that 
\be \la{sigma 1}
\sigma(\mbf{k}) = \fr{V P'}{v\, T} =\fr{\langle \mbf{k} | \mbf{k} \rangle P'}{2 \omega_k v  \langle\omega_k, \ell, m | \omega_k, \ell, m \rangle} =  \sum_{X_f}  \fr{ |\langle X_f; 0| S | X_i; \mbf{k}\rangle  |^2}{2 \omega_k v  \langle \omega_k, \ell, m | \omega_k, \ell, m \rangle}
\ee
In order to make contact with the process (\ref{process}), let us rewrite the term $|\langle X'; 0| S | X; \mbf{k}\rangle  |^2$ in the numerator using the completeness of the $| \omega, \ell, m \rangle$ states:
\ba
&& \!\!\!\!\!\!\!\!\!\!\!\!\!\!\!\!\!\!\!\!\!\! |\langle X_f; 0| S | X_i; \mbf{k}\rangle  |^2 = \nonumber \\
&=& \sum_{\ell, m, \ell', m'} \int \fr{d \omega}{2 \pi}  \fr{d \omega'}{2 \pi} \langle \mbf{k} | \omega, \ell, m \rangle  \langle X_i; \omega, \ell, m  | S^\dag | X_f; 0 \rangle \langle X_f; 0| S | X_i; \omega', \ell', m' \rangle \langle \omega', \ell', m'  | \mbf{k} \rangle \nonumber \\
&=&  \sum_{\ell, m, \ell', m'} \int \fr{d \omega}{2 \pi}  \fr{d \omega'}{2 \pi} \fr{2\omega_k v}{k^2} (2 \pi)^4 \delta(\omega_k - \omega) \delta(\omega_k - \omega')    \nonumber \\
&& \qquad \qquad \quad \times Y_\ell^m (\hat{\mbf{k}}) Y_{\ell'}^{m'*} (\hat{\mbf{k}})  \langle X_i; \omega, \ell, m  | S^\dag | X_f; 0 \rangle \langle X_f; 0| S | X_i; \omega', \ell', m' \rangle \nonumber \\
&=& \fr{2\omega_k v}{ k^2} \sum_{\ell, m, \ell', m'} (2 \pi)^2 Y_\ell^m (\hat{\mbf{k}}) Y_{\ell'}^{m'*} (\hat{\mbf{k}})  \langle X_i; \omega_k, \ell, m  | S^\dag | X_f; 0 \rangle \langle X_f; 0| S | X_i; \omega_k, \ell', m' \rangle  \, .  \la{sigma manip 1}
\ea
This expression simplifies significanly if instead of considering $\sigma(\mbf{k})$ we calculate its average over all possible directions of $\mbf k$:
\be
\sigma (k) \equiv \int \fr{d \Omega}{4 \pi} \, \sigma (\mbf{k}) \, .
\ee
In fact,
\ba
&& \int \fr{d \Omega}{4 \pi} \, \fr{2\omega_k v}{ k^2} \sum_{\ell, m, \ell', m'} (2 \pi)^2 Y_\ell^m (\hat{\mbf{k}}) Y_{\ell'}^{m'*} (\hat{\mbf{k}})  \langle X_i; \omega_k, \ell, m  | S^\dag | X_f; 0 \rangle \langle X_f; 0| S | X_i; \omega_k, \ell', m' \rangle = \nonumber \\ 
&& \qquad \qquad \qquad \qquad \quad  = \fr{2 \pi \omega_k v}{k^2} \sum_{\ell, m} |\langle X_f; 0| S | X_i; \omega_k, \ell, m \rangle |^2 \, ,  \la{sigma manip 2}
\ea
which means that
\be
\sigma (k) = \fr{\pi}{k^2} \sum_{l, m} P (\omega_k, \ell, m) \, .
\ee
This result must coincide with the cross section for the process (\ref{process}) averaged over all $m$'s and summed over all $\ell$'s. Therefore, we conclude that 
\be
\sigma (\omega, \ell, m) = \fr{ (2 \ell+1) \pi  P}{k^2} \, .
\ee
This equation is the analog of eq. (\ref{sigmak}). Notice that, even though this result was derived in the case of a spin 0 particle, it holds also in the case of higher spin particles, with the \emph{orbital} angular momentum $\ell$ replaced by the \emph{total} angular momentum $j$.

It is equally easy to derive the expression for the decay rate in eq. (\ref{decay rate l=1}). Once again, we will start by considering a process involving emission of a particle with definite momentum:
\be
X_i \to X_f + \mbf{k} \, .
\ee
If we denote with $P'$ the probability for one such process to occur regardless of the final state $X_f$ of the spinning object, then the decay rate is given by
\begin{eqnarray}
\Gamma = \l[ \fr{V}{(2 \pi)^3} \int {d^3 k} \r]    \fr{P'}{T} = \int \fr{d^3 k}{(2 \pi)^3} \sum_{X_f} \fr{ |\langle X_f; \mbf{k}| S | X_i; 0\rangle  |^2}{2 \omega_k  \langle \omega_k, \ell, m | \omega_k, \ell, m \rangle} \, ,
\end{eqnarray}
 where we have performed the same manipulations used to obtain eq. (\ref{sigma 1}). We can now rewrite the integral over $k$ in spherical coordinates and follow steps similar to those in eqs. (\ref{sigma manip 1}) and   (\ref{sigma manip 2}) (this time summing over all possible directions rather than averaging) to finally obtain:
\begin{eqnarray}
\fr{d \Gamma (\omega, \ell, m)}{d \omega} = \fr{k^2 P}{2 \pi v \omega^2} \, ,
\end{eqnarray}
where $P$ is now the probability for the process (\ref{Hawking-process}) summed over all possible final states $X_f$. Once again, this result is valid also for higher spin particles provided one replaces $\ell \to j$ \, .

\section{Spherical helicity states} \la{appd}

In order to extend our approach to superradiant scattering to particles with non-zero spin, it is convenient to introduce spherical helicity states. Usually, 1-particle states for a given mass $\mu$ and spin $s$ are classified by diagonalizing simultaneously the momentum operator $\mbf P$ and the helicity  $\mbf J \cdot \mbf P / | \mbf P| $. These states are denoted as $| \mbf k, \lambda \rangle $ and are the familiar plane wave states. However, one could also choose to diagonalize instead $H, J^2, J_z$ and $\mbf J \cdot \mbf P / | \mbf P| $, in which case the states would be denoted as $| \omega, j, m, \lambda \rangle$. These are known as \emph{spherical helicity states}~\cite{landau1971course}. By extending the normalization convention (\ref{normalization of spherical states}) for spin 0 particles, we will assume that these states are normalized as follows:
\be \la{ normalization of spherical helicity states}
\langle \omega', j', m', \lambda' |  \omega, j, m, \lambda \rangle = 2 \pi \delta (\omega - \omega') \delta_{jj'} \delta^{m m'} \delta_{\lambda \lambda'} \, .
\ee

In Section \ref{higher spins}, we need the matrix elements that connect these different basis of states. It is easy to realize that they must take the form
\begin{eqnarray}
\langle \mbf k, \lambda' | \omega, j, m, \lambda \rangle = 2 \pi \delta (\omega_k - \omega) \delta_{\lambda \lambda'} F (\mbf k, j, m, \lambda)  \, .
\end{eqnarray}
We will now determine the explicit form of the function $F (\mbf k, j, m, \lambda)$. Following~\cite{landau1971course} (albeit with slightly different conventions) we rewrite the momentum $\mbf k$ as a rotation $R$ acting the vector $ k \hat z$ pointing in the $z$-direction. At the level of the Hilbert space, this translates into the relation $| \mbf k, \lambda \rangle = U(R) |  k \hat z, \lambda \rangle$. Note that rotations do not change the helicity, since the latter is a scalar under rotations. Then, we have
\begin{eqnarray}
\langle \mbf k, \lambda | \omega, j, m, \lambda \rangle &=& \langle  k \hat z, \lambda | U^\dag (R) | \omega, j, m, \lambda \rangle \nonumber \\
&=& \sum_{j'm'\lambda'}\int_M^\infty \fr{d \omega}{2 \pi} \langle  k \hat z, \lambda | \omega', j', m', \lambda' \rangle \langle \omega', j', m', \lambda' | U^\dag (R)| \omega, j, m, \lambda \rangle \nonumber \\
&=& \sum_{m'} \langle  k \hat z, \lambda | \omega, j, m', \lambda \rangle  \langle  j, m' | U^\dag (R)|  j, m \rangle \nonumber \\
&= & \langle  k \hat z, \lambda | \omega, j, \lambda, \lambda \rangle  [D^{(j)}_{ m\lambda } (\hat k) ]^* \equiv 2 \pi \delta (\omega_k - \omega) f_{\lambda j} (\omega) [D^{(j)}_{ m\lambda } (\hat k) ]^* \, ,
\end{eqnarray}
where in the second line we used the completeness of states, in the third line the fact that energy, helicity, and total angular momentum are scalars under rotations and in the fourth line the fact that the helicity is, by definition, the component of the angular momentum in the direction of the momentum. We also introduced the Wigner rotation matrices $D^{(j)}_{m' m} (\hat k) $, which, as we can see, completely determine how the matrix elements depend on the direction of $\hat k$. In order to determine $f_{\lambda j} (\omega)$, we rewrite the normalization condition (\ref{ normalization of spherical helicity states}) using the completeness of the plane wave states:
\begin{eqnarray}
&& \langle \omega', j', m', \lambda' |  \omega, j, m, \lambda \rangle = \sum_{\lambda''} \int \fr{d^3 k}{(2\pi)^3}  \langle \omega', j', m', \lambda' | \mbf{k}, \lambda'' \rangle \langle \mbf{k}, \lambda'' | \omega, j, m, \lambda \rangle \nonumber \\
&& \qquad\qquad  = \int \fr{d \Omega \, d \omega \, k^2}{(2 \pi)^3 2 \omega v }  \delta_{\lambda \lambda'}  2\pi \delta(\omega_k - \omega') 2\pi \delta(\omega_k - \omega) f_{\lambda j'}^*(\omega')f_{\lambda j}(\omega)[D^{(j)}_{ m \lambda} (\hat k) ]^* D^{(j')}_{ m'\lambda} (\hat k) \nonumber \\
&& \qquad\qquad  = 2 \pi \delta (\omega - \omega') \delta_{\lambda \lambda'} \fr{k^2}{2 \omega v (2 \pi)^2} f_{\lambda j'}^*(\omega)f_{\lambda j}(\omega) \int d \Omega [D^{(j)}_{ m \lambda} (\hat k) ]^* D^{(j')}_{ m' \lambda} (\hat k) \nonumber \\ 
&& \qquad\qquad  = 2 \pi \delta (\omega - \omega') \delta_{jj'} \delta_{mm'} \delta_{\lambda \lambda'} \fr{k^2}{2 \omega v (2 \pi)^2} | f_{\lambda j}(\omega)|^2 \fr{4\pi}{2j + 1} \, .
\end{eqnarray}
By comparing the last line with eq. (\ref{ normalization of spherical helicity states}) we can easily determine $f_{\lambda j}(\omega)$. Finally, using the fact that the Wigner matrices are closely related to the spin-$s$ spherical harmonics $_s Y_{jm}$~\cite{Goldberg:1966uu}, we can write our matrix elements as follows:\footnote{Note that, in~\cite{Goldberg:1966uu}, the Wigner matrix $D^{(j)}_{m' m}$ describes a rotation that takes $\hat k$ into $\hat z$, whereas here we are adopting the opposite convention~\cite{Sakurai:2011zz}. Therefore, their $D^{(j)}_{m' m}$ corresponds to our $[D^{(j)}_{m m'}]^*$.}
\begin{eqnarray}
\langle \mbf k, \lambda' | \omega, j, m, \lambda \rangle= \delta_{\lambda \lambda'} \fr{\sqrt{2 \omega v}}{k}\,  (2 \pi)^2 \delta (\omega_k - \omega)  \,_{-\lambda} Y_{jm} (\hat{\mbf k}) \, .
\end{eqnarray}
This result is a direct generalization of the spin 0 result in eq. (\ref{k to omega l m }).

\section{Vector bound states with higher values of $\ell$ and $j$} \la{higher l bound states}

Suppressing all the index structure, there are in principle two possible classes of dissipative couplings which one could write down for a vector field. Schematically, they read
\begin{eqnarray} \la{possible vector couplings} 
(\d)^N \Phi_{0} (R)^N \mathcal{O} \quad \text{and} \quad (\d)^N \Phi_{I} (R)^{N+1} \mathcal{O}   \,  ,
\end{eqnarray}
where the partial derivatives $\d$ are all spatial derivatives (time derivatives would simply lead to subleading corrections).  The goal of this appendix is to determine which coupling yields the dominant contribution to the absorption of a vector bound state with a given set of quantum numbers $j$ and $\ell$. Notice that, according to the usual rules of addition of angular momentum, we can only have $j = \ell-1, \ell, \ell + 1$ for a vector field.

For scalars, we saw that dissipation of modes with orbital angular momentum $\ell$ is controlled at leading order by a coupling to a composite operator that involves $\ell$ spatial derivatives. For vectors, the relationship between the number of spatial derivatives and the orbital angular momentum $\ell$ is a bit more subtle. To untangle this relationship, let us first look closely at the constraint equation that determines the zero component of the vector field in terms of its spatial components. Using the ansatz for the mode functions given by Eqn.~(\ref{mode functions of spin 1 bound state}), we can solve the constraint $\nabla_\mu \Phi^\mu=0$ at lowest order in the  gravitational interaction to find 
\begin{eqnarray} \la{S(r)} 
S_{nj\ell}(r) \simeq \frac{1}{\mu}\sqrt{\frac{2\ell+1}{2j+1}}C_{1 \ell }(j,0; 0, 0)\left\{ \left[\frac{\d R_{n\ell}(r)}{\d r}-\ell \frac{R_{n\ell}(r)}{r}\right]+ (2\ell+1) \frac{R_{n\ell}(r)}{r} \delta^j_{\ell-1} \right\} . \,\,\,
\end{eqnarray}
There are a few points to notice about this expression. First, the Clebsch-Gordan coefficient $C_{1 \ell }(j,0; 0, 0)$, and therefore $S_{nj\ell}(r)$, is non-zero only when $j=\ell \pm1$. This means that the first class of couplings in (\ref{possible vector couplings}) cannot affect modes with $j = \ell$. Second, $S_{nj\ell}(r)$ has a different radial dependence for $j = \ell + 1$ and $j = \ell - 1$. In fact, because $R_{n\ell}(r)$ depends on $r$ schematically as
\be \label{R_nl scaling}
R_{n\ell}(r) \sim (G M \mu^2)^{3/2} \left[(G M \mu^2 r)^\ell + \mathcal{O}\left((G M \mu^2 r)^{\ell+1}\right) \right] \, ,
\ee
we see that the term in square brackets in Eq. (\ref{S(r)}) vanishes to leading order in $r$. Therefore, for small values of $r$ and $(G M \mu) \ll 1$, we find that 
\begin{subequations} \la{S scaling}
\begin{eqnarray} 
S_{nj\ell}(r) &\sim &\frac{(GM\mu^2)^{5/2}}{\mu}(GM\mu^2r)^{\ell} \quad \quad \quad\text{for $j=\ell +1$} \\
\la{S(r) scaling} 
&\sim &\frac{(GM\mu^2)^{5/2}}{\mu}(GM\mu^2r)^{\ell-1} \quad \quad \text{for $j=\ell -1$} \,  .
\end{eqnarray}
\end{subequations}
Let us now consider separately the three possible cases $j= \ell$, $j = \ell + 1$ and $j = \ell - 1$.

In the $j=\ell$ case we have that $S_{nj\ell}=0$, and thus we must consider a coupling involving~$\Phi_I$. Based on the $r$ dependence of $R_{n \ell}$ shown in (\ref{R_nl scaling}), it is clear that we need (at least) $\ell$ derivatives acting on~$\Phi^I$ in order for the absorption and emission rates not to vanish when we set $r = 0$. Therefore, the most relevant coupling for modes with $j = \ell$ is schematically of the form
\begin{eqnarray} \la{j = l vector coupling}
(\d)^\ell \Phi_{I} (R)^{\ell+1} \mathcal{O} \qquad \qquad \text{for $j=\ell$} \, ,
\end{eqnarray}
where the operator $\mathcal{O}$ is in the spin-$j$ representation of the rotation group. This means that, for instance, dissipation of the mode with $j = \ell = 1$ will be determined by the interaction $\epsilon_{IJK} \d^M \Phi^N R_M{}^I R_N{}^J \mathcal{O}^K$.

The story becomes slightly more complicated when $j=\ell +1$, as now both kinds of couplings shown in (\ref{possible vector couplings}) are in principle allowed.  Once again, we can determine the minimum number of derivatives needed by examining the small-$r$ limit of $R_{n\ell}$ and $S_{nj\ell}$. Based on Eqs. (\ref{R_nl scaling}) and (\ref{S(r) scaling}), we see that for $j=\ell +1$ we need (at least) $\ell$ derivatives both for the $\Phi_0$ coupling and the $\Phi_I$ coupling in order to get a non-zero contribution to the absorption rate. However, by comparing these two couplings we find that 
\be
(\d)^\ell \Phi_{0} (R)^\ell \mathcal{O}  \sim (G M \mu) \times (\d)^\ell \Phi_{I} (R)^{\ell+1} \mathcal{O} \qquad \qquad  \text{for $j=\ell+1$} \, .
\ee
Consequently, the leading dissipative coupling for the $j=\ell +1$ mode in the limit $G M \mu \ll~1$ is the one that involves the spatial components $\Phi_I$, just as in the $j=\ell$ case. Notice however that now the operator $\mathcal{O}$ has $j = \ell +1$ indices, i.e. one more index compared  to the operator appearing in the coupling (\ref{j = l vector coupling}). This means that, for example, the leading interactions for the mode with $j = 2$ and $\ell = 1$ are $\mu \, \d^I \Phi^J R_I{}^K R_J{}^L \mathcal{O}_{KL}$ and $ \d^I E^J R_I{}^K R_J{}^L \mathcal{O}_{KL} \supset  \d^I \d^0 \Phi^J R_I{}^K R_J{}^L \mathcal{O}_{KL}$ with $\mathcal{O}_{KL}$ symmetric and traceless. The factor of $\mu$ in the first interaction ensures that it vanishes in the massless case, as it should since it's not gauge invariant. These two interactions are then of the same order, because $\d_0 \Phi^I \sim \mu \Phi^I $.

Let us finally consider the $j=\ell-1$ case. According to Eq.~(\ref{S(r) scaling}) we have $S_{nj\ell} \sim r^{\ell-1}$, and therefore the two  dissipative channels we should in principle consider are now $(\d)^{\ell-1} \Phi_{0} (R)^{\ell-1} \mathcal{O}$ and $(\d)^\ell \Phi_{I} (R)^{\ell+1} \mathcal{O}$, where the operator $\mathcal O$ is in the spin-$j$ representation in both cases. Notice in particular the different number of spatial derivatives. When we compare these two terms, we find that 
\be
(\d)^{\ell-1} \Phi_{0} (R)^{\ell-1} \mathcal{O}  \sim (G M \mu)^{-1} \times (\d)^\ell \Phi_{I} (R)^{\ell+1} \mathcal{O} \qquad \qquad \text{for $j=\ell-1$} \, .
\ee
Therefore, we see that the leading interaction for modes with $j = \ell -1$ is the one that involves $\Phi_0$. This means in particular that dissipation of the mode with $j=1$ and $\ell =2$ will be determined by the coupling $ E_I R^I{}_J \mathcal O^J \supset \d_I \Phi_0 R^I{}_J\mathcal O^J$. Notice that a coupling of the form $\d_I \Phi_0 R^I{}_J\mathcal O^J$ just by itself would be suppressed by an extra power of $\mu$ because it's not gauge invariant.

Now that we have determined what is the most relevant interaction for a given value of $j$ and $\ell$, estimating the difference between absorption and emission rates is merely a matter of dimensional analysis. Following the same steps adopted in Section~\ref{bound states}, and using the small-$r$ expressions in Eqs. (\ref{R_nl scaling}) and (\ref{S scaling}), we find that for $j = \ell$,
\begin{eqnarray}
\Delta\Gamma \propto |\d^\ell R_{n \ell} |^2 \, \fr{\rho (\mu - m \Omega)}{\mu} \propto (G M \mu^2)^{3+ 2\ell} \, \fr{(GM)^{2+2\ell}(\mu - m \Omega)}{\mu} \propto (G M \mu)^{5 + 4 \ell} (\mu - m \Omega) , \nonumber
\end{eqnarray}
with $-j \leqslant m \leqslant j$. Notice that in the second step, the factor of $(GM)^{2+2\ell}$ comes from the dimensions of the dissipative coefficient $\gamma$. For $j = \ell +1$ the power counting is very similar, except that there is an extra factor of $\mu^2$ coming from the interaction squared, with an additional $(GM)^2$ thrown in for good measure to comply with basic dimensional analysis.  The end result is therefore
\begin{eqnarray}
\Delta \Gamma \propto (GM \mu)^2 \times (G M \mu)^{5 + 4 \ell} (\mu - m \Omega) = (G M \mu)^{7 + 4 \ell} (\mu - m \Omega)
\end{eqnarray}
It is easy to check that this expression reduces to the result we found in Sec. \ref{vector bound states} when $\ell = 0$. Finally, for modes with $j = \ell -1$, we have instead 
\begin{eqnarray}
\Delta\Gamma \propto |\d^{\ell-1} S_{n j \ell} |^2 \, \fr{\rho (\mu - m \Omega)}{\mu} \propto \fr{(G M \mu^2)^{3+ 2\ell}}{\mu^2} \, \fr{(GM)^{2\ell}(\mu - m \Omega)}{\mu} \propto (G M \mu)^{3 + 4 \ell} (\mu - m \Omega) , \nonumber
\end{eqnarray}
with $-j \leqslant m \leqslant j$. Thus we see that, for any given $\ell$, the difference between absorption and emission rates for the mode with the lowest possible $j$ is larger than that for the modes with higher $j$.

As we already mentioned in Sec. \ref{vector bound states}, our results are compatible with those of~\cite{Rosa:2011my,Pani:2012bp}. In those papers, the authors considered the limit $\Omega \to 0$ rather than the one $\Omega \ll \mu$ that we have have used in Sec. \ref{vector bound states}. If we set $\Omega = 0$ in the result above, it is easy to see that they can be all combined to give:
\begin{eqnarray}
\Delta\Gamma \propto  \mu (G M \mu)^{5 + 2 \ell + 2j},
\end{eqnarray}
for the superradiant modes (i.~e.~ for $j>0$) in perfect agreement with~\cite{Rosa:2011my,Pani:2012bp} (see footnote on page \pageref{footnote notation} for a comment about the different notation used in those papers).


\bibliography{biblio-no-fermions}{}
\bibliographystyle{hunsrt}

\end{document}